\documentclass[aps,prx,superscriptaddress,showpacs,keywords,reprint]{revtex4-1}
\usepackage{graphicx,color}
\usepackage{amsfonts,amsmath, amsthm, amssymb,graphicx,hyperref}
\usepackage{stackengine}
\usepackage{xcolor}

\stackMath
\newcommand{\be}{\begin{equation}}

\newcommand{\ee}{\end{equation}}
\newcommand{\ben}{\begin{eqnarray}}
\newcommand{\een}{\end{eqnarray}}
\newcommand{\bes}{\begin{subequations}}
\newcommand{\ees}{\end{subequations}}
\newcommand{\bF}{\begin{figure}}
\newcommand{\eF}{\end{figure}}

\def\tr{ {\rm{Tr }}\,}

\newcommand{\kt}{\rangle}
\newcommand{\br}{\langle}

\usepackage{lineno,hyperref}
\modulolinenumbers[5]

\begin{document}
\title{Quantum signatures of chaos, thermalization and tunneling in the exactly solvable few body kicked top}
\author{Shruti Dogra}
\affiliation{Department of Physics, Indian Institute of Technology Madras, Chennai, India 600036}
\email{shrutidogra.iiserm@gmail.com}
\email{vmadhok@gmail.com}
\email{arul@iitm.ac.in}

\author{Vaibhav Madhok}
\affiliation{Department of Physics, Indian Institute of Technology Madras, Chennai, India 600036}

\author{Arul Lakshminarayan}
\affiliation{Department of Physics, Indian Institute of Technology Madras, Chennai, India 600036}

\begin{abstract}
Exactly solvable models that exhibit quantum signatures of classical chaos are both rare as well as important - more so in view of the fact that the mechanisms for ergodic behavior and thermalization in isolated quantum systems and its connections to non-integrability are under active investigation. In this work, we study quantum systems of few qubits collectively modeled as a kicked top, a textbook example of quantum chaos. In particular, we show that the 3 and 4 qubit cases are exactly solvable and yet, interestingly, can display signatures of ergodicity and thermalization. Deriving analytical expressions for entanglement entropy and concurrence, we see agreement in certain parameter regimes between long-time average values and ensemble averages of random states with permutation symmetry. Comparing with results using the data of a recent transmons based experiment realizing the 3-qubit case, we find agreement for short times, including a peculiar step-like behaviour in correlations of some states. In the case of 4-qubits we point to a precursor of dynamical tunneling between what in the classical limit would be two stable islands. Numerical results for larger number of qubits show the emergence of the classical limit including signatures of a bifurcation.
\end{abstract}
\maketitle
\section{Introduction}
In a modest pursuit of the esthetic attributed to the probabilist Feller that ``the best consists of the general embodied in the concrete" \cite{Billingsley}, we consider extreme
quantum cases of the kicked top, a widely studied text-book model of quantum chaos \cite{Haake,Peres02,KusScharfHaake1987,KusMostowskiHaake1988,Zyczkowski1990,Gerwinski1995,Wang2004, LombardiMatzkin2011,Ghose, mrgi14, Lewenstein-arxiv-2018,Bhosale-pre-2017}, which has also been implemented in experiments \cite{Chaudhary,Neill16}. The general issues at hand are the emergence
of classical chaos from a linear quantum substratum and, more recently, the role of quantum chaos in the thermodynamics of closed quantum systems \cite{CassidyEtal2009,SantosRigol2010,Rigol16}. Vigorous progress is being made in studying thermalization of isolated quantum systems that could be either time-independent or periodically forced \cite{JensenShankar1985,Deutsch91,Srednicki94,Rigol2009,CassidyEtal2009,SantosRigol2010,CiracHastings2011,
DeutchLiSharma2013,LangenEtal2013,Rigol16,LucaRigol2014,
LazaDasMoess2014,LazDasMoess2014pre,Haldar2018,Neill16,Kaufman2016,ClosEtal2016,HazzardEtal2014}. 
  Entanglement within many-body states in such quantum chaotic systems 
drives subsystems to thermalization although the full
state remains pure and of zero entropy, see \cite{Kaufman2016} for a demonstration with cold atoms.

Quantum chaos \cite{Gutzwiller1990,Haake} and, consequently, eigenstate thermalization hypothesis \cite{Srednicki94,Rigol16} enables one to use individual states for ensemble averages. For periodically driven systems that do not even conserve energy, a structureless ``infinite-temperature" ensemble emerges in strongly non-integrable regimes \cite{LucaRigol2014,LazDasMoess2014pre}. 
A recent 3-qubit experiment, using superconducting Josephson junctions, that simulated the kicked top \cite{Neill16} (see also \cite{Madhok2018_corr}) purported to remarkably demonstrate such a thermalization. Although such behavior has been attributed to non-integrability \cite{Neill16,Rigol16}, we exactly solve this 3-qubit kicked top and also point out that it can be interpreted as a special case of an {\it integrable} model, the well-known transverse field Ising model. Interestingly, we also solve the 4-qubit case exactly, where there is no such evident connection to an already known integrable model.

The Arnold-Liouville notion of integrability requires sufficient number of independent constants of motion in involution.
It is well-known that in finite dimensional quantum systems this notion can be debated, wherein any system is integrable 
as the projectors on eigenstates form a set of independent mutually commuting quantities, for example see \cite{YusShastry2013}.
However, in this work, we use integrability more in the sense of the traditional definition of the existence of constants that arise from symmetries and 
whose forms are independent of the parameters of the system. This is a pragmatic approach and in line with current understanding that would
classify the nearest neighbor transverse field Ising model as integrable and one with an additional longitudinal field, or a transverse field Ising model with nearest as well as next-nearest neighbor interactions as non-integrable.

Nonintegrable, chaotic, systems may be solvable in some tangible sense, the textbook examples
of the tent map and the bakers map are solvable, despite being completely chaotic. The Arnold cat map, and its
quantizations also admit analytical solutions despite being hyperbolic and chaotic. Nevertheless, this is very rare, and 
restricted to abstract models. No known model that has a mixed phase space, with both regular and chaotic orbits, is also known to be exactly solvable in 
the same sense. Attempts at constructing such models include the piecewise linear ``lazy bakers map". 
 The kicked top, in the limit of an infinite number of qubits displays a standard transition to Hamiltonian chaos, including a mixed phase space, and it is remarkable that many of the features are already reflected in the solvable few qubit cases as we show in this paper.

For example, we obtain explicit formulas for entanglements generated for the 3 and 4-qubit cases and the compare the former with data from the experiment in \cite{Neill16} and find very good agreement. The infinite time average of single qubit entanglement is found analytically for some initial states and at a special and 
large value of the forcing, for all initially unentangled coherent states. These are shown to tend to that obtained from relevant (random matrix) 
ensembles, in some cases even exactly coinciding with them and  thus displaying thermalization. These demonstrate that even in the deep quantum regime, the transition to what in the classical limit becomes chaos is reflected in the time-averaged entanglement. While the connections between chaos and entanglement in the semiclassical regime is now well studied \cite{MillerSarkar,Wang2004, LombardiMatzkin2011,Ghose, trail2008entanglement, Lewenstein-arxiv-2018,Lakshminarayan,BandyopadhyayArul2002,Bandyopadhyay04,
ScottCaves2003,Ghose,Lakshminarayan16}, such systems are typically not analytically tractable and appeal is made to statistical modeling based on random matrix theory.
Remarkably, there are interesting quantum effects in the few-body systems we study here. 
We find the presence of dynamical tunneling 
\cite{DavisHeller1981,LinBallentine1990,Peres1991,Tomsovic98b,SrihariBook} 
between what appears in the classical limit as symmetric regular regions.
This results in extremely slow convergence of subsystem entropies in the near-integrable regime that happens for some states of the 4-qubit case.
In the near-integrable regime the exactly calculable tunneling splitting is shown to result in this long-time dynamics. The kicked-top experiment involving the spin of cold Cs atoms has already observed such tunneling \cite{Chaudhary} but our observations provide a connection between 
the number of qubits and a system parameter at which such tunneling occurs. This may open windows to study the interplay of chaos and tunneling even in systems having a small number of qubits.
 
 \subsection{The model \label{Theory}}
 \begin{figure}
 \centering
 \includegraphics[scale=1]{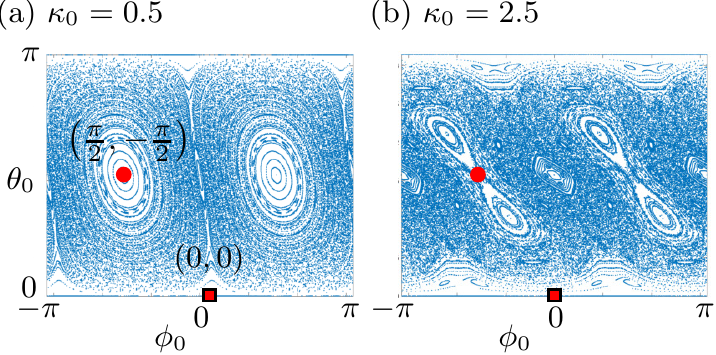}
 \caption{(a) Regular and (b) mixed phase space structures 
 resulting from the classical chaotic dynamics. Points labelled 
 with red square and red circle correspond to initial states $\Theta=0, \Phi=0$
 on a period-4 orbit
 and  $\Theta=\pi/2, \Phi=-\pi/2$ at the centre of regular island respectively.}
 \label{fig:classical}
\end{figure}
The quantum kicked top is a combination of a rotation and a torsion, the Hamiltonian \cite{KusScharfHaake1987,Haake,Peres02} is given by
\begin{equation}
\label{Eq:QKT}
H=\frac{\kappa_0}{2j}{J_z}^2 \sum_{n = -\infty}^{ \infty} \delta(t-n\tau)+\frac{p}{\tau} \, {J_y}.
\end{equation}
Here $J_{x,y,z}$ are components of the angular momentum operator $\mathbf{J}$. 
The time between periodic kicks is $\tau$. The Floquet map is the unitary operator,
\begin{equation}
 \mathcal{U} = \exp\left [-i (\kappa_0/2j \hbar) J_z^2 \right]\exp\left[-i (p/\hbar) J_y\right],
\end{equation}
which evolves states just after a kick to just after the next. The parameter $p$ 
measures rotation about the $y$ axis, and in the following we set $\hbar=1$ and $p=\pi/2$.
The  parameter $\kappa_0$, which is the magnitude of a 
twist applied between kicks controls the transition and measure of chaos. If it vanishes, the 
dynamics is simply a rotation. As the magnitude of the total angular momentum is conserved, 
the quantum number $j$, with eigenvalues of $\mathbf{J}^2$ being $j(j+1)\hbar^2$,  
is a good one. The classical limit, when $j \rightarrow \infty$ is a map of the unit 
sphere phase space $X^2+Y^2+Z^2=1$ onto itself with the variables 
being $X,Y,Z=J_{x,y,z}/j$ 
and is given by (at $i^{\textrm{th}}$ iteration of the map)
\begin{eqnarray}
 X_{i}&=&Z_{i-1}\cos(\kappa_0 X_{i-1})+Y_{i-1}\sin (\kappa_0 X_{i-1}),\nonumber \\
 Y_{i}&=&-Z_{i-1}\sin(\kappa_0 X_{i-1})+Y_{i-1}\cos (\kappa_0 X_{i-1}),\nonumber \\
 Z_{i}&=&-X_{i-1}.
\end{eqnarray}

Numerical iterations for various different initial conditions:
$(X_{0}, Y_{0}, Z_{0})$, and for two 
strengths of the chaos, $\kappa_0=0.5$ and $2.5$, are shown in Fig.~(\ref{fig:classical}).
These display what may be termed as regular and mixed phase space structures respectively, with the measure
of chaotic oribits at $\kappa_0$ being negligibly small.
For $\kappa_0=0$ the classical map is evidently integrable, being just a rotation, but for $\kappa_0>0$ chaotic orbits appear in the phase space and when  $\kappa_0>6$ it is essentially fully chaotic. Connection to a many-body model can be made by considering the large $\mathbf{J}$ spin as the total spin 
of spin=1/2 qubits, replacing $J_{x,y,z}$ with $\sum_{l=1}^{2j} \sigma^{x,y,z}_l/2$ \cite{Milburn99,Wang2004}. The Floquet operator is then that of $2j$ qubits, an Ising model with all-to-all homogeneous coupling and a transverse magnetic field:
\begin{equation}
\label{uni}
{\mathcal U}=\exp\left(-i \frac{\kappa_0}{4j}  \sum_{ l< l'=1}^{2j} \sigma^z_{l} \sigma^z_{l'}\right)
\exp\left( -i \frac{\pi}{4} \sum_{l=1}^{2j}\sigma^y_l \right).
\end{equation}
Here $\sigma^{x,y,z}_l$ are the standard Pauli matrices, and an overall phase is neglected. 
In general only the $2j+1$ dimensional permutation symmetric subspace of the full $2^{2j}$ dimensional space is relevant to the kicked top.

Note that for $\kappa_0$ that are multiples of $2 \pi j$, $\mathcal{U}$ is a local operator and does not create entanglement, we therefore restrict attention to the interval $\kappa_0 \in [0, \pi j]$.
The case of 2-qubits, $j=1$, has been analyzed in \cite{RuebeckArjendu2017} wherein interesting arguments have been proposed for the observation of structures not linked to the classical limit. In this case, several quantum correlation measures were also calculated in \cite{Bhosale-PRE-2018}.
For $j=3/2$, the three qubit case, as all-to-all is just nearest neighbor with periodic boundary conditions, it is a nearest neighbor kicked transverse Ising model, known to be integrable \cite{Prosen2000,ArulSub2005}. The Jordan-Wigner transformation renders it a model of noninteracting fermions that 
can be immediately solved. This is also the case that was considered in the superconducting Josephson junction experiment \cite{Neill16} that treated it as chaotic. For higher values of the spin $j$, the model maybe considered few-body realizations of non-integrable systems. 

In the following we will mostly be studying time evolution from initial states that are localized in 
the spherical phase space, and these are the standard $SU(2)$ coherent states. Permutation symmetric 
initial states used are coherent states located at 
\begin{eqnarray}
 X_{0}&=&\sin\theta_0 \cos\phi_0, \nonumber \\ 
 Y_{0}&=&\sin \theta_0 \sin\phi_0, \nonumber \\ 
 Z_{0}&=&\cos \theta_0,
 \end{eqnarray}
on the phase space sphere and given by~\cite{Glauber,Puri},
\begin{equation}
 |\theta_0,\phi_0\kt = \otimes^{2j} 
(\cos(\theta_0/2) |0\kt + e^{-i \phi_0} \sin(\theta_0/2) |1\kt).
\end{equation}

\section{Analytical solution of the three-qubit case}
From Eq.~(\ref{uni}), the unitary Floquet operator for $2j=3$-qubits,
that simulate the dynamics of a spin-$3/2$ under a kicked top Hamiltonian is given by,
\begin{eqnarray}
 \label{eq1a}
 \mathcal{U} = \exp && \left({-i \frac{\kappa_0}{6} (\sigma_1^z\sigma_2^z+\sigma_2^z\sigma_3^z+\sigma_3^z\sigma_1^z)} \right).
 \nonumber \\
 && \exp \left({-i \frac{\pi}{4}(\sigma_1^y+\sigma_2^y+\sigma_3^y)} \right), 
\end{eqnarray}
where all the terms have their usual meanings as defined in Section~\ref{Theory}.
The solution to the 3-qubit case proceeds from the general observation that 
\[ [\mathcal{U},\otimes_{l=1}^{2j} \sigma^y_l]=0, \] {\it i.e.,} there is
an ``up-down" or parity symmetry.

\begin{figure}
 \centering
 \includegraphics[scale=1,keepaspectratio=true]{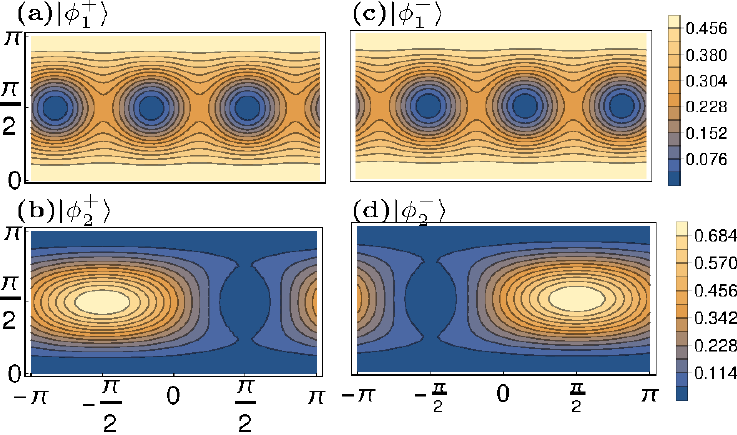}
 \caption{Husimi (quasiprobability distribution, $|\langle \phi_i|\theta_0,\phi_0 \rangle|^2$) plots
 for a set of four three-qubit bases states ($|\phi_i\rangle$), where $|\theta_0,\phi_0\rangle$ is 
 an arbitrary three-qubit, parametrized by ($\theta_0,\phi_0$).  \label{husimi3q}}
\end{figure}

The standard 4-dimensional spin quartet permutation symmetric space with $j=3/2$,
$\{|000\kt, |W\kt=(|001\kt+|010\kt+|100\kt)/\sqrt{3},
|\overline{W}\kt =(|110\kt+|101\kt+|011\kt)/\sqrt{3},|111\kt\}$
is parity symmetry adapted to form the basis
\begin{eqnarray}
|\phi^{\pm}_1\kt&=&\frac{1}{\sqrt{2}}(|000\kt \mp i | 111 \kt), \\
|\phi_2^{\pm}\kt&=&\frac{1}{\sqrt{2}} (|W\rangle \pm i |\overline{W}\kt).
\end{eqnarray}
These are parity eigenstates such that $\otimes_{l=1}^{3} \sigma^y_l|\phi_j^{\pm}\kt=\pm |\phi_j^{\pm}\kt$.
Notations employed reflect the usage of $|W\kt$ as the standard $W-$ state
of quantum information and the $|\phi^{\pm}_1\kt$ correspond to the standard
GHZ states. To visualize these basis states the contour plots of their quasiprobability distribution in the phase space
is shown in Fig.~(\ref{husimi3q}). We see that while the GHZ class of states are localized prominently at the poles 
of the sphere, the superposition of the $W$ states are localized at the equatorial plane and peak at $(\theta_0=\pi/2,\phi_0=\pm \pi/2)$.
Interestingly these points correspond to low-order periodic points for the classical map and form the most important initial states
to evolve for the quantum system.
In this basis, the unitary operator $\mathcal{U}$ is given by
\begin{equation} 
\label{eq6}
 \mathcal{U} = \begin{pmatrix}
            \mathcal{U}_{+} & 0 \\ 0 & \mathcal{U}_{-}
            \end{pmatrix}, 
\end{equation}
where $0$ is a $2 \times 2$ null matrix,
and $2\times2$-dimensional blocks $\mathcal{U}_{+}$ ($\mathcal{U}_{-}$) are written 
the bases $\{ \phi_{1}^{+}, \phi_{2}^{+} \}$ ($\{ \phi_{1}^{-}, \phi_{2}^{-} \}$),
are in the  positive (negative)-parity subspaces respectively. Explicitly, these have matrix
elements
\begin{equation}
  \label{eq:Uplusm}
  \mathcal{U}_{\pm} = \pm  e^{\mp \frac{i \pi}{4}} e^{-i \kappa} \begin{pmatrix}
   \frac{i}{2}e^{-2i \kappa} & \mp \frac{\sqrt{3} }{2} e^{-2i \kappa} \\
   \pm \frac{\sqrt{3}}{2} e^{2i \kappa} &  -\frac{i}{2}e^{2i \kappa}
   \end{pmatrix}.
 \end{equation}
For simplicity the parameter $\kappa=\kappa_0/6$ is used in these expressions. 
Expressing $\mathcal{U}_+$ as a rotation $e^{-i \gamma \vec{\sigma} \cdot\hat{\eta}}$
by angle $\gamma$ about an axis 
$\hat{\eta}=\sin{\theta} \cos{\phi} \, \hat{x}
 + \sin{\theta} \sin{\phi} \, \hat{y} + \cos{\theta} \, \hat{z}$,
upto a phase. 
On comparison with Eq.~(\ref{eq:Uplusm}), we obtain,
$ \cos{\gamma} =\frac{1}{2} \sin{2\kappa}$, 
 $\phi=\pi/2 +2 \kappa$, and 
 $\sin{\theta} \sin{\gamma} = \sqrt{3}/2$.
 To evolve initial states we need $\mathcal{U}^n$ and therefore
$\mathcal{U}_{\pm}^n$, which is 
 explicitly given by,
\begin{equation}
\label{eq:Upluspowern}
\mathcal{U}_{\pm}^n = (\pm 1)^n e^{-i n (\pm  \frac{\pi}{4}+\kappa)}
\begin{pmatrix}
 \alpha_n &
   \mp \beta_n^* \\
   \pm \beta_n &  
   \alpha_n^*
 \end{pmatrix}, 
\end{equation}   
   where,
   \begin{eqnarray}
   \label{eq:alphabetan}
    \alpha_n &=& T_n(\chi)+\frac{i}{2}\, U_{n-1}(\chi) \cos 2\kappa \quad \textrm{and}\\
    \label{eq:betan}
   \beta_{n} &=& (\sqrt{3}/2)\, U_{n-1}(\chi) \,e^{2 i \kappa}.
   \end{eqnarray}
The Chebyshev polynomials $T_n(\chi)$ and $U_{n-1}(\chi)$ are defined as $T_n(\chi)=\cos(n \gamma)$ and $U_{n-1}(\chi)=\sin(n \gamma)/\sin \gamma$ \cite{mason2002chebyshev} with 
   $\chi=\cos{\gamma}=\sin(2\kappa)/2$. Also note that $|\alpha_n|^2+|\beta_n|^2=1$. This follows both from the unitarity of $\mathcal{U}_{\pm}$ as well
   as a polynomial Pell identity satisfied by the Chebyshev polynomials, namely
   \begin{equation}
   T^2_n(x)+(1-x^2) U_{n-1}^2(x)=1.
   \end{equation}
Remarkably, one can also view this as a new proof of the Pell identity satisfied by Chebyshev polynomials through the unitarity
of quantum mechanics.

Note also that the range of $\chi$ is restricted in this case to $|\chi|\leq 1/2$, which in addition to the
general identity $|T_n(\chi)|\leq 1$, 
also implies that $|U_{n-1}(\chi)|\leq 2/\sqrt{3},$ which follows from Eq.~(\ref{eq:betan}).

It is now straightforward to do time evolution, 
for an arbitrary three-qubit permutation symmetric state,
and thereafter study its various properties.
We further analyse two widely different 
three-qubit states ((i) $|0,0\kt$ and (ii) $|\pi/2, -\pi/2\kt$) in detail.
For these two states, we obtain the exact expressions for 
linear entropy of a single-party reduced density matrix, time-average
of the linear entropy, and  
concurrence between any two qubits as 
a measure of entanglement.
These analytical expressions are verified numerically and also 
compared, where possible, with the data from the superconducting 
transmon qubits experiment of \cite{Neill16}. We particularly 
considered these two examples due to their preferential behaviors 
as classical phase space structures.
A three-qubit state $\otimes^3|0\kt$ corresponds to 
coherent state at $|0,0\kt$ which is on the period-4 orbit
whose classical correspondence is shown with a square
in Fig.~(\ref{fig:classical}), while $\otimes^3|+\kt_y$ corresponds 
to the coherent state at $|\pi/2,-\pi/2\kt$,
 which is a fixed point on the classical phase space. This becomes unstable as we move 
 from regular to mixed phase space at $\kappa_0=2$ and is indicated by a circle in Fig.~(\ref{fig:classical}).
\subsection{Initial state $|000\rangle=|\theta_0=0,\phi_0=0\kt$ \label{example1}}
Let us consider the state on the period-4 orbit, corresponding to the 
coherent state at $|0,0\kt$ which is $\otimes^3|0\kt$.
\begin{equation}
\label{eq29}
\begin{split}
|\psi_n\kt &=  \mathcal{U}^n |000\rangle =
  \frac{1}{\sqrt{2}} \mathcal{U}^n \left( |\phi_{1}^{+} \rangle + |\phi_{1}^{-} \rangle \right) \\
& = \frac{1}{\sqrt{2}} \left( \mathcal{U}_{+}^n |\phi_{1}^{+} \rangle + \mathcal{U}_{-}^n |\phi_{1}^{-} \rangle \right)\\
& =\frac{1}{2}e^{-i n \left(\frac{3 \pi}{4}+\kappa\right)}\left\lbrace (1+i^n) \left(
 \alpha_n |000\rangle + i \beta_n |\overline{W} \rangle
 \right) \right.   \\ &  \left. + (1-i^n) \left( i \alpha_n |111\rangle - \beta_n |W \rangle
 \right) \right\rbrace. 
 \end{split} 
 \end{equation}
From this the $1$ and $2$ qubit reduced density matrices $\rho_1(n)=\text{tr}_{2,3} (|\psi_n \rangle \langle \psi_n |)$,
 $\rho_{12}(n)=\text{tr}_{3}( |\psi_n \rangle \langle \psi_n |)$ are obtained. 
 The entanglement of one qubit with the other
 two is found as the linear entropy $1-\tr[\rho_1(n)^2]$, and from the $2$-qubit reduced matrix, the entanglement 
 between two qubits is found as the concurrence \cite{Wootters}.
 \subsubsection{The linear entropy}
 It turns out that for even values of the time $n$, say $n=2m$, $\rho_1(2m)$ is diagonal, 
 whose diagonal elements are,
 $\lambda(2m,\kappa_0)$ and $1-\lambda(2m, \kappa_0)$, from which the linear entropy,
\begin{equation} \label{ent3q}
 S_{(0,0)}^{(3)}(2m,\kappa_0)=2 \lambda(2m,\kappa_0)(1- \lambda(2m,\kappa_0)),
\end{equation}
where the eigenvalue,
\begin{equation}
\label{eq:lamb000}
 \lambda(2m,\kappa_0) = \frac{1}{2} U^{2}_{2m-1}(\chi) =\frac{2}{3}|\beta_{2m}|^2.
\end{equation}

For odd values of $n$, $\rho_1(n)$ is not diagonal, but a peculiar result is obtained.
One can evolve the even $n=2m$ states one step backward 
in time
\begin{equation}
 |\phi_{2m-1}\rangle=\mathcal{U}^{-1} |\phi_{2m}\rangle,
\end{equation}
where $\mathcal{U}$ is the Floquet operator in Eq.~(\ref{eq1a}).
Let $m$ itself be an even integer, which implies that only the first half of the state
in Eq.~(\ref{eq29}) survives. Then upto an overall phase,
using the nonlocal part of the unitary operator $\mathcal{U}$,
the state upto local unitary operations is 
\begin{eqnarray}
  |\phi_{2m-1}\rangle &\stackunder{=}{\text{loc}}&  
  e^{i \kappa (\sigma_1^z\sigma_2^z+\sigma_2^z\sigma_3^z+\sigma_3^z\sigma_1^z)} 
  \left(
 \alpha_{2m} |000\rangle + i \beta_{2m} |\overline{W} \rangle
 \right), \nonumber \\
  &=& e^{3i \kappa} \alpha_{2m} |000\rangle + i e^{-i \kappa} \beta_{2m} |\overline{W} \rangle, \nonumber \\
 &=&  \mathcal{V}\otimes \mathcal{V}\otimes \mathcal{V} |\phi_{2m} \rangle,
\end{eqnarray}
where single qubit unitary operator $\mathcal{V} = e^{i \kappa \sigma_z}$.
Thus the three qubit state $|\psi_{2m-1}\rangle$, after odd 
numbered implementations of the unitary operator $\mathcal{U}$ are 
local unitarily equivalent to the state obtained after $2m$
implementations of $\mathcal{U}$
and hence all entanglement properties including entropy and concurrence
are left unchanged for an odd-to-even time step. A similar 
situation holds when $m$ is odd.
Therefore for a pair of consecutive implementations, entanglement among the 
qubits does not change, giving rise to step like features in the variation of entropy 
and concurrence with time. In particular
\begin{equation}
\label{eq:entsteps}
S_{(0,0)}^{(3)}(2n-1,\kappa_0)= S_{(0,0)}^{(3)}(2n,\kappa_0), \;\; n=1,2,\cdots.
\end{equation}
 \begin{figure}
 \centering
 \includegraphics[scale=1]{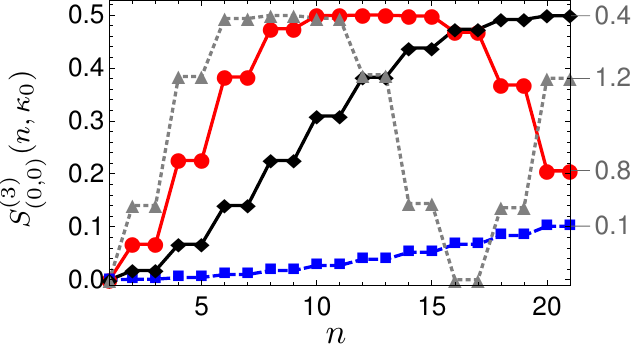}
 \caption{Linear entropy of a single qubit reduced state versus $n$ is plotted
 for initial state $|000\kt$ at different values of $\kappa_0=0.1, \; 0.4, \; 0.8$ and $1.2$ as 
 labelled on the right end of each curve.}
 \label{fig:entropy3qubit1}
\end{figure}
 This step like feature in the variation of entropy is illustrated for a few values of $\kappa_0$ in
 Fig.~(\ref{fig:entropy3qubit1}). It is seen that there is a monotonic increase of the initial 
 rate of entropy production as a function of $\kappa_0$. This gives way to non-monotonic behavior both in
 time and in the parameter $\kappa_0$. The initial rate can be simply quantified by the entanglement entropy
 at $n=1$. Again using the linear entropy we have as a special case that 
\begin{equation}
\label{eq:000time1}
 S_{(0,0)}^{(3)}(1,\kappa_0)=\sin^2(\kappa_0/3)\left(1-\frac{1}{2}\sin^2(\kappa_0/3)\right),
\end{equation}  
which increases monotonically till $\kappa_0=3 \pi/2$ where acquires the maximum value of $1/2$ which is also the upper-bound. We will see that the case of $\kappa_0=3 \pi/2$ is one of maximal chaos in some sense for $j=3/2$.

{For small $\kappa_0$, the growth of the entropy is 
$S_{(0,0)}^{(3)}(1,\kappa_0) \approx \kappa_0^2/9$. From Fig.~(\ref{fig:entropy3qubit1}) it is seen that even for small
values of $\kappa_0$ the entropy eventually becomes large and the maximum allowed value of $1/2$ is reached. As the classical
dynamics for small $\kappa_0$ is regular, the large value of the entanglement reached is intriguing. We now estimate the time it takes for the entanglement to reach nearly the maximum value.
The state in Eq.~(\ref{eq29})
clearly distinguishes times modulo 4. If the time $n$ is odd and $\beta_n$ vanishes (the conditions under which this happens is discussed below), the resultant state is 
the GHZ one with an equal superposition of $|000\kt$ and $|111\kt$ which is such that the reduced density matrices are maximally mixed
and hence have maximum entropy. If the time $n$ is even and $\beta_n$ vanishes, there is no entanglement as the state becomes a tensor product, this also being
apparent from the Eqs.~(\ref{ent3q}) and ~(\ref{eq:lamb000}).}

From Eq.~(\ref{eq:betan}), the vanishing of $\beta_n$ corresponds to the zeros of the Chebyshev polynomials of the second kind, $U_{n-1}(\chi)$, which are at $\chi=\chi_k=\cos(\pi k/n)$ and $k=1,2, \cdots,n-1$. Thus we are looking for values of $n$ such that 
\begin{equation}
\frac{1}{2}\sin(\kappa_0/3)=\cos(\pi k /n),
\end{equation} 
which may be found from the continued fraction convergents of 
$r=\cos^{-1}[ \sin(\kappa_0/3)/2]/\pi$. For small $\kappa_0$ ($\ll 1$), 
$r \lesssim 1/2$	 the first non-zero convergent is $1/2$ and therefore the 
second is of the form $a_1/(2 a_1+1)$ where $a_1$ is an integer. Identifying this 
with $k/n$ we see that $n$ is an odd integer and hence this corresponds to the case of 
maximum, or at least near-maximum, entanglement. Taylor expanding the $\sin$ and the $\cos^{-1}$ and
retaining the lowest order terms then gives an estimate of the time $n_*$ at which the entanglement, for the first time, reaches nearly the maximum as
\begin{equation}
\label{eq:000maxtime}
n_* \approx 2 \left[ \dfrac{3 \pi}{2 \kappa_0}-\frac{1}{2}\right] +1 \approx \left[ \frac{3 \pi}{\kappa_0}\right],
\end{equation} 
and the time at which it gets unentangled, for the first time, is $\sim 2 n_*$. We see from Fig.~(\ref{fig:entropy3qubit1}) 
that these are excellent estimates even when $\kappa_0$ is as large as $0.4$ or $0.8$.

{The formation of non-classical states such as the GHZ in this instance is a forerunner
of dynamical tunneling as for small $\kappa_0$ the islands at the ``poles" of the phase space
sphere can start to localize states for large values of $j$. This effect is seen prominently in the long-time
averages. The intriguing increase of entanglement with time, even for small $\kappa_0$ in these states
therefore has very different origins than the non-integrability of the kicked top. 
}
\subsubsection{Long time averaged linear entropy}

The infinite time average of the linear entropy, which 
can be easily obtained from Eq.~(\ref{ent3q}), maybe inaccessible experimentally
but is of definite interest from the point of view of thermalization and it also
is a way to study the influence of the parameter $\kappa_0$ directly.
 We need to use only even values of the time as for this state due to the
 property discussed above. We have
\begin{eqnarray} \label{ent4}
 S^{(3)}_{(0,0)}(2m,\kappa_0) &=& U^{2}_{2m-1}(\chi)-\frac{1}{2} U^{4}_{2m-1}(\chi) \\
 &=& \frac{\sin^2 2m \gamma}{\sin^2\gamma} -\frac{1}{2} \frac{\sin^4 2m \gamma}{\sin^4\gamma}.
\end{eqnarray}
The time-averaged linear entropy is thus given by
\begin{eqnarray} \label{ent5}
 \br S^{(3)}_{(0,0)}(\kappa_0) \kt &=& \lim_{N \rightarrow \infty}\frac{1}{N} \sum_{m=0}^{N-1} S^{(3)}_{(0,0)}(2m, \kappa) \\
 &=& \frac{1}{2\sin^2\gamma}  -\frac{3}{16\sin^4\gamma},
\end{eqnarray}
where we have used that $\br \sin^2(2m \gamma)\kt =1/2$ and $\br \sin^4(2m \gamma)\kt =3/8$,
assuming that $\gamma \neq 0,\pi/2, \pi$.
Further, using $\cos \gamma = \frac{1}{2}\sin 2\kappa= \frac{1}{2}\sin (\kappa_0/3)$, we obtain
the average explicitly in terms of $\kappa_0$ as
  \begin{equation} \label{eq:avgpiby2_1}
  \br S^{(3)}_{(0,0)}(\kappa_0) \kt=\frac{5-2\sin^2(\kappa_0/3)}{\left(4-\sin^2(\kappa_0/3)\right)^2}, \, 0<\kappa_0<3 \pi.
 \end{equation}
This attains its maximum value of $1/3$ at $\kappa_0=3\pi/2$. This may be used as a probe 
to understand the process of thermalization, which is discussed later in this section.
However it is appropriate to point out that $\br S^{(3)}_{(0,0)}(\kappa_0) \kt$ is 
discontinuous at $\kappa_0=0$ as it vanishes at $\kappa_0=0$ but is $5/16$ for arbitrarily
small and nonzero values. Thus in this deep quantum regime, the state that starts off 
from the period-4 orbit gets entangled to a large extent even when the oribit is classically 
stable. However this is reflected in the infinite time average which includes highly nonclassical
time scales, as discussed above.

\subsubsection{Concurrence}

While the linear entropy is a measure of entanglement of one qubit with the other two,
the entanglement between any two qubits is quantified by the concurrence. Due to the
permutation symmetry in the state it does not matter which two qubits are considered,
there is only one concurrence. The concurrence is derived from the two-qubit reduced 
density matrix, as opposed to the entanglement of one qubit which needs only the one-qubit
state. If $\rho_{12}$ is the two-qubit state, then its concurrence is given by
\begin{equation}
\label{eq:conc_defn}
\mathcal{C}(\rho_{12})=\text{max}\left(0, \sqrt{\lambda_1}-\sqrt{\lambda_2}-\sqrt{\lambda_3}-\sqrt{\lambda_4} \right),
\end{equation} 
where $\lambda_i$ are eigenvalues in decreasing order of $(\sigma_y \otimes \sigma_y)\rho_{12} (\sigma_y \otimes \sigma_y) \rho_{12}^*$,
where $\rho_{!2}^*$ is conjugation is in the standard ($\sigma_z$) basis.

 \begin{figure}[h!]
 \centering
 \includegraphics[scale=1]{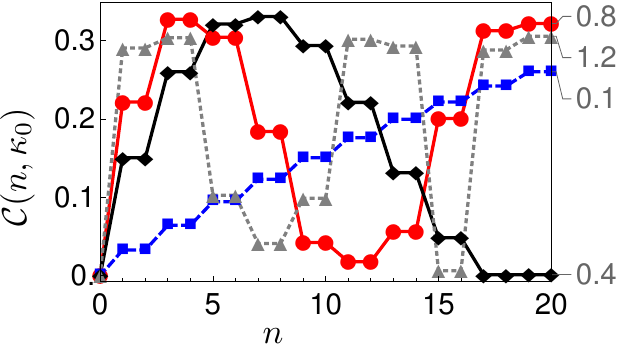}
 \caption{Concurrence of a two-qubit reduced state versus $n$ is plotted
 for $\kappa_0=0.1, \; 0.4, \; 0.8$ and $1.2$ as labelled on the right end of each curve.}
 \label{fig:conc3qubit1}
\end{figure}

An exact expression for concurrence amongst any two qubits in the state $|\psi_n\kt$ of 
Eq.~(\ref{eq29}) is possible to obtain explicitly as
the two-qubit state is an ``$X$ state" \cite{YuEberly2007} when the time $n$ is even.
A two-qubit reduced density operator of $\rho_{12}(n)$ obtained 
by tracing out one of the qubits in $|\psi_n\kt\br \psi_n |$ is given by,
\begin{equation}
 \rho_{12}(n)= \begin{pmatrix}
 |\alpha_n|^2 & 0 & 0 & -\frac{i}{\sqrt{3}} \alpha_n \beta_n^{*} \\
 0 & \frac{1}{3} |\beta_n|^2 & \frac{1}{3} |\beta_n|^2 & 0 \\
  0 & \frac{1}{3} |\beta_n|^2 & \frac{1}{3} |\beta_n|^2 & 0 \\
\frac{i}{\sqrt{3}} \alpha_n^{*} \beta_n & 0 & 0 & \frac{1}{3}|\beta_n|^2
 \end{pmatrix},
\end{equation}
whose concurrence is found from the general formula for the $X$ states \cite{YuEberly2007},
\begin{equation}
\label{eq:concstate000}
\begin{split}
&\mathcal{C}(n,\kappa_0)= \\ & 2\,  \text{max} \left[ 0, \frac{1}{3}|\beta_{n}|^{2} - \frac{1}{\sqrt{3}} |\alpha_{n}||\beta_{n}|,
  -( \frac{1}{3}|\beta|_{n}^{2} - \frac{1}{\sqrt{3}} |\alpha_{n}||\beta_{n}|)\right]\\ 
&=2 \left| \frac{1}{3} \vert \beta_n \vert^2
 -\frac{1}{\sqrt{3}} \vert \alpha_n \vert \vert \beta_n \vert \right|\\
&= \left| U_{n-1}(\chi) \right| \left| \frac{1}{2} 
 \vert U_{n-1}(\chi) \vert- \sqrt{1-\frac{3}{4} \vert U_{n-1}(\chi) \vert^2 } \right|,
\end{split}
\end{equation}
where we recall for convenience that $\chi=\cos \gamma=\sin(2\kappa)/2=\sin(\kappa_0/3)/2$.
This is valid when the time $n$ is even,but from the arguments presented in the discussion
of the entanglement entropy it follows that 
\begin{equation}
\label{eq:concsteps}
\mathcal{C}(2m-1,\kappa_0)=\mathcal{C}(2m,\kappa_0), \, m=1,2,\cdots.
\end{equation}

See Fig.~(\ref{fig:conc3qubit1}) for the variation of the concurrence with time for the same 
values of $\kappa_0$ as used in the previous figure. As with the case of the linear entropy, the 
concurrence initially increases monotonically with $\kappa_0$ as well as with time. Once again it 
is of interest to see how much concurrence is produced in simply the first step and this is 
\begin{equation}
\mathcal{C}(1,\kappa_0)=\sin(\kappa_0/3)\left[\sqrt{1-\frac{3}{4}\sin^2(\kappa_0/3)} - \frac{1}{2}\sin(\kappa_0/3)\right],
\end{equation}
which is valid when $0 \le \kappa_0 \le 3 \pi$, and beyond this the concurrence is periodic.
Interestingly this is monotonic in $\kappa_0$ only till $\kappa_0=\pi/2$, where it attains the 
maximum value of $(\sqrt{13}-1)/8 \approx 0.3257$. This is in contrast to the linear entropy or entanglement 
of one qubit with the rest which grows till $\kappa_0=\pi$.

 \begin{figure}[h!]
 \centering
 \includegraphics[scale=1]{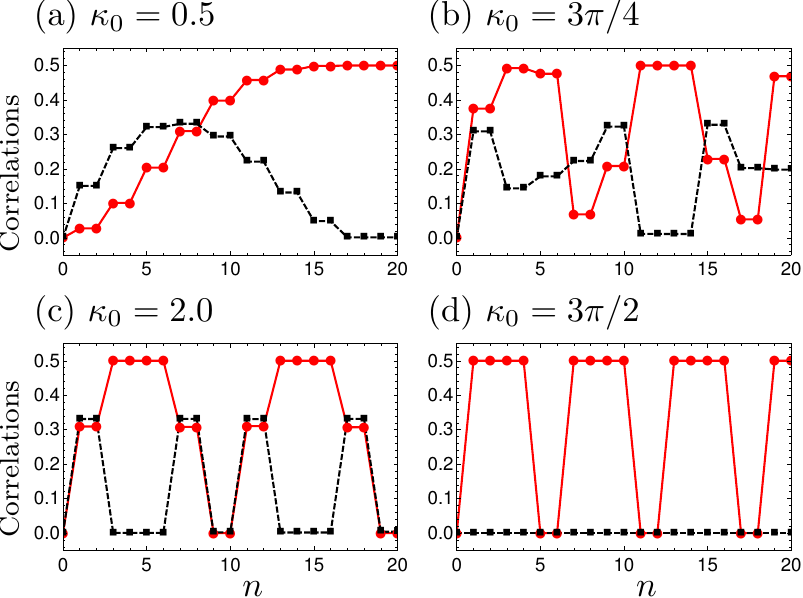}
 \caption{Solid curve with circles and dashed curve with squares 
 show the variation of entropy of a single qubit reduced state and concurrence between a pair
 of two qubits respectively, with $n$ as three-qubit initial state $|000\kt$ evolves under $\mathcal{U}^n$. 
 Parts (a), (b), (c), and (d) correspond to different values of chaoticity parameter ($\kappa_0$)
 as mentioned.}
 \label{fig:entconc}
\end{figure}
It is useful to compare the concurrence and entanglement entropy directly and this is
illustrated in Fig.~(\ref{fig:entconc}) where for $4$ value of $\kappa_0$ these are 
plotted as a function of time. It is seen that while initially both of them grow,
after a certain time, the concurrence starts to decrease while the entanglement continues
to increase. This is the phase where entanglement is started to be shared globally rather
than in bipartite manner. In this case of only $3$ qubits, this implies that tripartite entanglement 
starts to significantly grow after this time. It is also seen that when the entanglement entropy 
is the maximum possible, concurrence is at a minimum, and sometimes vanishes. This is consistent with the
fact that entanglement is monogamous and hence cannot be simultaneously shared among the three qubits. 
It is interesting that the simple formulas derived for this system illustrates these more general 
features. In particular it is clear from Eq.~(\ref{ent3q}) and Eq.~(\ref{eq:concstate000}) that while
both the entanglement and concurrence vanish when $U_{n-1}(\chi)=0$, the concurrence also vanishes when 
$U_{n-1}(\chi)=\pm 1$, a case that corresponds to a maximum entanglement.
More discussion on this is also found in \cite{Madhok2018_corr}.

A curious case is obtained when $\kappa_0=3 \pi/2$ when $\cos\gamma=1/2$ and hence $\gamma =\pi/3$ and
$U_{n-1}(\chi)=\sin(2 \pi n/3)/\sin(\pi/3)$, which takes the value $0$ when $n \,(\text{mod}\,3)=0$,
is $+1$ when $n \,(\text{mod}\,3)=1$ and is $-1$ when $n \,(\text{mod}\,3)=2$. This implies that when  $n \,(\text{mod}\,6) \neq 0$ or $-1$
the entanglement entropy is the maximum possible value of $1/2$ while the concurrence vanishes for all values of time
$n$, as seen in the last panel of Fig.~(\ref{fig:entconc}). Thus in this case the entanglement is shared only in a tripartite 
manner. We will return to this case later, but note here that indeed special values of such parameters in 
Floquet spin systems display similar behavior with large multipartite entanglement \cite{SunilMishraArulSubhra}.

\subsection{Initial state $|+++\rangle_y=|\theta_0=\pi/2, \phi_0=-\pi/2\kt$ and beyond\label{example2}}
We considered in some detail the fate of the state $|000\kt$, we now 
study the case of the three-qubit state $|\psi_0\rangle=|+++\rangle_y$, where
$|+\rangle_y=\frac{1}{\sqrt{2}}(|0\rangle+i|1\rangle)$
is an eigenvector of $\sigma_y$ with eigenvalue $+1$.
The former is an eigenstate of the interaction term in the Floquet operator 
$\mathcal{U}$, while the latter is the eigenstate of the field.
When $|+++\rangle_y$ is the initial state, it's evolution lies entirely in the 
positive parity sector as it can be also written as
$\otimes^3|+\kt_y=(|\phi_1^+\kt +\sqrt{3} i |\phi_2^+\kt)/2$. 
As a coherent state it corresponds to being localized at $|\pi/2,-\pi/2\kt$. 
The corresponding classical object is a fixed point that is stable till $\kappa_0=2$.
The time evolved state is then 
\begin{equation}
|\psi_n\kt=\mathcal{U}^n|+++\kt_y= e^{-in(\frac{\pi}{4}+\kappa)}\left( \gamma_n |\phi_1^{+} \rangle + \delta_n |\phi_2^{+} \rangle \right),
\label{eq:fixedptstate}
\end{equation}
where $\gamma_n=(\alpha_n-i\sqrt{3}\beta_n^{*})/2$ and $\delta_n=(\beta_n+i\sqrt{3}\alpha_n^{*})/2$.
and the $\alpha_n$ and $\beta_n$ are same as in Eq.~(\ref{eq:alphabetan}).

One can obtain the single-party reduced state by tracing out any two-qubits,  $\rho_{1}(n) =$
\begin{eqnarray}
&& \begin{pmatrix}
           \frac{1}{2}  
           & -\frac{i}{3} \left( |\delta_n|^2 - \sqrt{3}\, \text{Im}( \gamma_n \delta_n^{*} ) \right) \\
           \frac{i}{3} \left( |\delta_n|^2 - \sqrt{3}\, \text{Im}( \gamma_n \delta_n^{*} ) \right) &
           \frac{1}{2}  
            \end{pmatrix}.  \nonumber \\ \label{eq-e2-5}
\end{eqnarray}
The eigenvalues of $\rho_1(n)$ are simple and given by 
$2\chi^2U^2_{n-1}(\chi)$ and $1-2\chi^2U_{n-1}^2(\chi)$; hence the linear entropy is
\begin{equation}   
\label{eq-e2-8}
 S_{(\frac{\pi}{2}, -\frac{\pi}{2})}^{(3)}(n,\kappa_0)= 4\chi^2U^2_{n-1}(\chi) \left( 1-2\chi^2 U_{n-1}^2(\chi) \right).
\end{equation}
 \begin{figure}
 \centering
 \includegraphics[scale=1]{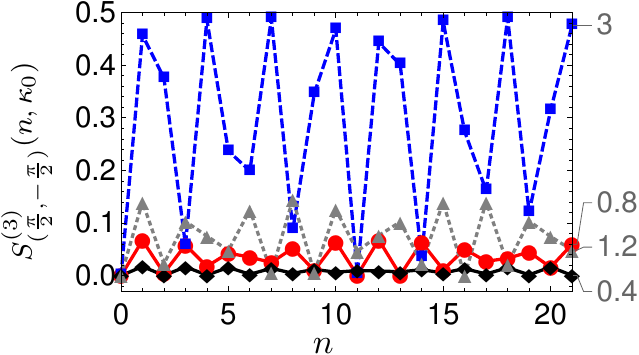}
 \caption{Linear entropy of a single qubit reduced state versus $n$ is plotted
 for different values of $\kappa_0$. Curves correspond to
 $\kappa_0=0.4, \; 0.8, \; 1.2$ and $3.0$ are shown by 
 solid line with diamonds, solid line with circles, dashed line with 
 triangles, and dashed line with squares respectively.}
 \label{fig:entropy3qubit2}
\end{figure}
Figure~(\ref{fig:entropy3qubit2}) shows the growth of the entanglement entropy in this state
as a function of time $n$ for four different values of $\kappa_0$. Comparing with Fig.~(\ref{fig:entropy3qubit1}) we
see that the entanglement increases much more slowly, in keeping with the classical 
interpretation of this state as being localized on a fixed point. However the initial $n=1$ is the same in 
both the cases, $S^{(3)}_{(\frac{\pi}{2},-\frac{\pi}{2})}(1,\kappa_0)$ is still given Eq.~(\ref{eq:000time1}),
and hence the entanglement after the first step is $\sim \kappa_0^2/9$ for small $\kappa_0$.

{A difference is seen at $n=2$ when 
\begin{equation}   
\label{eq:ppptime2}
 S_{(\frac{\pi}{2}, -\frac{\pi}{2})}^{(3)}(2,\kappa_0)= \sin^4(\kappa_0/3)\left(1-\frac{1}{2}\sin^4(\kappa_0/3)\right),
\end{equation}
thus while $S_{(0, 0)}^{(3)}(2,\kappa_0)=S_{(0, 0)}^{(3)}(1,\kappa_0)$, $S_{(\frac{\pi}{2}, -\frac{\pi}{2})}^{(3)}(2,\kappa_0)<S_{(\frac{\pi}{2}, -\frac{\pi}{2})}^{(3)}(1,\kappa_0).$ In fact the contrast with the state $|000\kt$ is most apparent when we observe that 
Eq.~(\ref{eq-e2-8}) implies that 
\begin{equation}
\label{eq:pppbound}
 S_{(\frac{\pi}{2}, -\frac{\pi}{2})}^{(3)}(n,\kappa_0)\leq 4 \chi^2 U_{n-1}^2(\chi) \leq \frac{4}{3}\sin^2(\kappa_0/3),
\end{equation}
the last inequality is due to the upper-bound $|U_{n-1}(\chi)|\leq 2/\sqrt{3}$ which as has been observed above holds 
due to the restriction $|\chi|\leq 1/2$. This inequality is useful for small $\kappa_0$ in which case we have
that $S_{(\frac{\pi}{2}, -\frac{\pi}{2})}^{(3)}(n,\kappa_0) \leq 4 \kappa_0^2/27$, and is hence very close to the
entanglement produced at the very first step, namely $\kappa_0^2/9$, and in particular has no secular growth towards
large entanglement.}

The  long-time average value of the linear entropy is calculated exactly as the case when the initial state
was $|000\kt$, and we therefore merely display the result
 \begin{equation} \label{eq:avgpiby2_2}
  \br S^{(3)}_{(\frac{\pi}{2},-\frac{\pi}{2})}(\kappa_0) \kt=\frac{\sin^2(\kappa_0/3)}{\left(4-\sin^2(\kappa_0/3)\right)^2}\left( 8-5\sin^2(\kappa_0/3)\right).
 \end{equation}
The major difference between the two initial states considered so far is apparent in this formula, as it is smooth
at $\kappa_0=0$ and vanishes at $\kappa_0=0$, unlike the Eq.~(\ref{eq:avgpiby2_1}). The fact that the classical orbit in this 
case is a fixed point as opposed to a period-4 orbit is notable. In the case of the state $|000\kt$, extremely non-classical
states such as the GHZ can form at sufficiently long times and leads to the large average. We will see that the state centered at the
fixed point can also have a large nonzero average for the case of $4$-qubits, again due to the formation of highly non-classical 
states mediated by tunneling.
 
\begin{figure}
 \centering
 \includegraphics[scale=1,keepaspectratio=true]{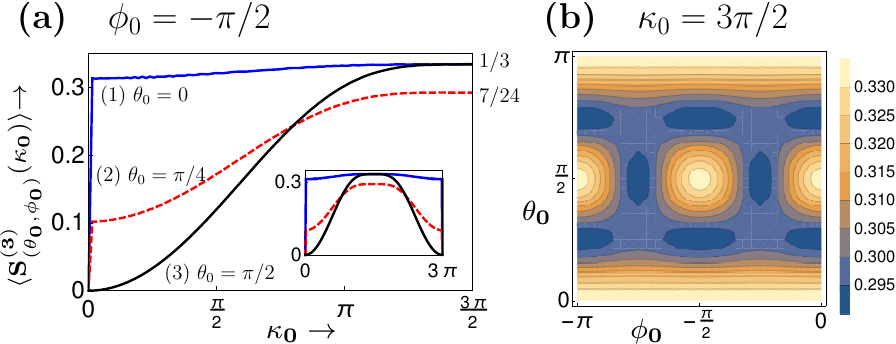}
 \caption{(a) Time averaged linear entropy, obtained over $n=1000$ periods,
 of a single qubit {\it vs} the parameter $\kappa_0$, for three initial coherent states $|\theta_0, \phi_0\kt$. The Eqs.~(\ref{eq:avgpiby2_1},~\ref{eq:avgpiby2_2}) apply to the curves labeled (1) and (3), as for $\theta_0=0$ the value of $\phi_0$ is immaterial on the sphere.  Inset shows the entanglement periodicity in the parameter at $\kappa_0=3\pi$. Part (b) displays the time averaged linear entropy across all initial coherent states for the value $\kappa_0=3\pi/2$ and is described by Eq.~(\ref{eq:entthetaphi}).}
\label{fig:entropyplot}
\end{figure}

Figure~(\ref{fig:entropyplot}) shows the long-time average  $\br S^{(3)}_{(\theta_0,\phi_0)}(\kappa_0)\kt$ as a function of $\kappa_0$, for the case of 3-qubits, and three initial states, two of them being what we just discussed, namely $|000\kt$ and $|+++\kt_y$, which correspond to the cases with $\theta_0=0$ and $\pi/2$ respectively. That they are in some sense
extreme cases is seen clearly in this figure. Each is seen to increase with the torsion $\kappa_0$ to $1/3$, while a state with $\theta_0=\pi/4$ (and in all cases $\phi_0=-\pi/2$) grows to $7/24$ which we will see is the lowest for any state. 
 The average value of the linear entropy in the $N$-qubit permutation symmetric subspace~\cite{SheshadriPRE2018} is given by  
\begin{equation}
 S_{RMT}(N)=\frac{N-1}{2N},
\end{equation}
and for $N=3$ this also gives $1/3$. For at least three particular initial states, with important classical phase space correspondences, $|0,0\kt\equiv |000\kt$ and $|\pi/2, \pm \pi/2\kt \equiv |\pm\pm\pm\kt_y$ this value is, remarkably, exactly attained for $\kappa_0=3\pi/2$, as easily verified from Eq.~(\ref{eq:avgpiby2_1}) and Eq.~(\ref{eq:avgpiby2_2}).
Thus ergodicity is attained as time averaged linear entropy approaches the state space averaged linear entropy.

Note that the $j=\infty$, classical system shows a transition to chaos in the same range of the parameter. 
While $j=3/2$ is too small to see effects such as the fixed points' loss of stability, the overall region surrounding the classical fixed points $(\theta_0,\phi_0)=(\pi/2, \pm \pi/2)$ being stable for small $\kappa_0$ and gradually losing stability as the parameter is increased is reflected in the gradual increase of average entropy corresponding to the
initial states $|\pi/2,\pm \pi/2\kt$ starting from $0$ when $\kappa_0=0$. Notice that from a purely quantum mechanical view, $\otimes^{2j} |\pm \kt_y$ are eigenstates of $\mathcal{U}$ at $\kappa_0=0$.  In contrast, the initial state  $|000\kt$ corresponds to a classical period-4 orbit and assumes entanglement entropy as large as $5/16$ for arbitrarily small $\kappa_0$. 
  
 \subsubsection{Arbitrary initial states, $\kappa_0=3 \pi/2$}
For the 3-qubit case, the case of $\kappa_0=3\pi/2$ is an extreme one, and the eigenvalues of $\mathcal{U}$
in this case are $\exp(\pm 2 \pi i/3)$ and $\pm \exp(\pm \pi i/6)$, implying that $\mathcal{U}^{12}=I$. Thus infinite 
time averages are finite ones over a period, in fact entanglement has a period of $6$ in this case and 
for arbitrary initial coherent states, the time-averaged entanglement entropy is obtained via a straightforward, if long, computation whose details 
we skip and state the result as 
\begin{equation}
\begin{split}
\br S^{(3)}_{(\theta_0,\phi_0)}(3\pi/2)\kt = &\frac{1}{48}[ 15+\cos(4 \theta_0) + \\ &(1+3 \cos(2 \theta_0)) \sin^4 \theta_0 \sin^2(2 \phi_0)].
\end{split}
\label{eq:entthetaphi}
\end{equation}
This takes values in the narrow interval $[7/24,1/3]$, and is shown in Fig.~(\ref{fig:entropyplot}). The minimum corresponds to several initial states including $|\pi/4,\pm \pi/2\kt$ and the maximum includes the $|0,0\kt$ and $|\pi/2, \pm \pi/2\kt$ states as already noted above. The structures seen are not directly linked to classical phase space orbits, except through shared symmetries \cite{RuebeckArjendu2017}, and cannot be expected to do so as the classical limit is for fixed $\kappa_0$ and $j \rightarrow \infty$. 
Nevertheless these results lend quantitative credence to thermalization in the sense that the time averaged entropy of subsystems of most states are close to the ensemble average for suitable large $\kappa_0$, even for the 3-qubit case \cite{Neill16,Rigol16}.which, when $\kappa_0=3\pi/2$ approaches $1/3$. 
Coincidentaly, as mentioned above, this is same as the average linear entropy 
of a single qubit reduced state in a set of random symmetric
three-qubit states. The calculations for the case of a general initial state and more figures of the long-time averages
are presented in the Appendix.
\section{Comparison with an experiment}

We analyse the data from a recent experiment~\cite{Neill16}, 
that demonstrates the kicked top dynamics of a spin-$3/2$,
using three superconducting transmon qubits. Experimental 
data corresponds to the two special initial states: $|0,0\kt$ and $|\pi/2,-\pi/2\kt$,
(whose analytical solutions are given in sections~\ref{example1} 
and \ref{example2} respectively), each one for two values of 
chaoticity parameter, $\kappa_0=0.5$ and $2.5$. 
Three-qubit state is experimentally initialized in given initial states (respectively),
and then allowed to undergo a series of kicks and evolutions, 
separately for $\kappa_0=0.5$ and $2.5$, as described
in~\cite{Neill16} for a total of $20$ time steps. 

Details of the analysis of the raw experimental data, which often has negative states,
is outlined. We analysed the complete quantum state tomographic 
data, obtained at the end of each time step.
State of a three-qubit system is obtained via complete quantum state 
tomography using a set of 64 projective measurements. These projective 
measurements are constructed by taking the combinations of Pauli-$x,y,z$ 
matrices ($\sigma_x$, $\sigma_y$ $\sigma_z$) and the Identity 
operator ($I$)~\cite{james-pra-2001, Neill16}.
These measurements are experimentally realized by various
single qubit rotations ($\mathcal{R}$) followed by $\sigma_z$ measurements
on individual qubits, that effectively performs a $\sigma_{i'}$
measurement (for $i'=x$, $\mathcal{R}=$Hadamard operator $(H_d)$; 
$i'=y$, $\mathcal{R}=$ Phase shift $(\mathcal{S}).H_{d}$; $i'=z$, 
$\mathcal{R}=I$)~\cite{Neill16}.
Multiple implementations of each measurement, provides the  
relative occupancy of the eight basis states of a three-qubit system.
 The resulting relative populations ($p_m$) of these eight
 states are thus obtained experimentally. 
 
 In order to 
compensate the effect of errors induced by the measurements,
 the intrinsic populations ($p_{int}$) are obtained
 via a correction matrix ($F$)~\cite{steffen-science-2006, lucero-prl-2008}. 
 We have, $p_{int}=F^{-1} p_m$, where $F=F_1 \otimes F_2 \otimes F_3$.
 $F_i$ is the measurement error corresponding to $i^{th}$ qubit,
 given as,
 \begin{displaymath}
  F_i= \left( \begin{array}{cc}
        f_0^{(i)} & 1-f_1^{(i)} \\ 1-f_0^{(i)} & f_1^{(i)}
       \end{array} \right).
 \end{displaymath}
Here, $f_0^{(i)}$ is the probability by which a state $|0\rangle$ 
of the $i^{th}$ qubit is correctly identified as $|0\rangle$, while 
$1-f_1^{(i)}$ is the probability by which,
a state that is actually $|0\rangle$ is being wrongly considered as $|1\rangle$.
$f_0^{(i)}$ and $f_1^{(i)}$ are termed as the measurement fidelities
of the basis states $|0\rangle$ and $|1\rangle$ respectively of the $i^{th}$ qubit.
Using a part of the measurement
data corresponding to the initial state preparation, we estimated the 
measurement fidelities as  $f_0^{(1)}=0.98$, $f_1^{(1)}=0.92$, $f_0^{(2)}=0.98$, 
$f_1^{(2)}=0.94$, $f_0^{(3)}=0.96$, $f_1^{(3)}=0.87$.
The intrinsic populations obtained in this manner are positive
(as observed till second decimal place). Using these intrinsic
population values, three-qubit density operators are obtained, 
that further undergo the convex optimization. The fidelities 
between the theoretically expected ($\rho_t$) and the experimentally
obtained ($\rho_e$) states is given by~\cite{Neill16}
\begin{equation}
 \mathcal{F}=Tr \sqrt{\sqrt{\rho_t}\rho_e\sqrt{\rho_t}}.
\end{equation}
These experimentally obtained three-qubit density operators are 
then used in our study to obtain the correlations, such as 
linear entropy of a single qubit reduced state and a two-qubit 
entanglement measure, concurrence.


Experimental data has its own imperfections, and the three-qubit experimental 
state may be not be permutation symmetric under qubit exchange. Therefore, corresponding to each three-qubit density operator, 
three single qubit reduced density operators (say $\rho_1$, $\rho_2$, $\rho_3$) and three 
two-qubit reduced density operators (say $\rho_{12}$, $\rho_{23}$, $\rho_{13}$) are obtained.
At each time step, using various single qubit and two-qubit density operators correlations such as linear entropy 
and concurrence are calculated respectively and their average behavior is observed.

\begin{figure}
 \centering
 \includegraphics[scale=1]{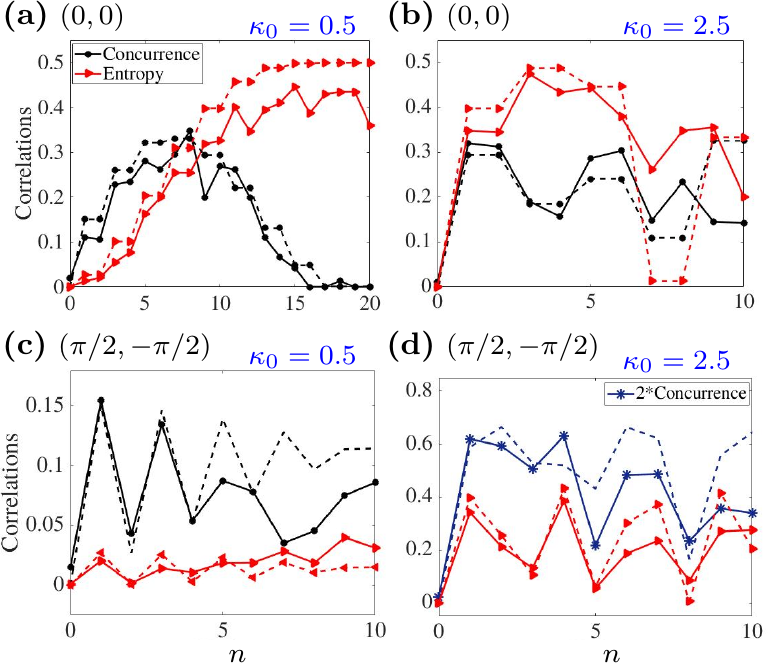}
 \caption{Plots showing analytical (dashed curves with markers), experimental 
 (solid curves with markers) and numerical (dashed) curves of linear entropy 
 and concurrence as a function of the number of kicks, as the initial state $|\psi_0\rangle$ is evolved  
  under repeated applications of operator $\mathcal{U}$.
 Parameters of the initial state, $(\theta_0, \phi_0)$, and 
 chaoticity parameter, $\kappa_0$, are specified in each figure.
 Analytical (wherever plotted) and numerical curves exactly overlap,
 and hence can not be seen separately.}
 \label{compk1tf4}
 \label{compk1tf1}
\end{figure}

Figure~\ref{compk1tf1} shows the comparison between the analytical results (dashed curves with markers) and those using 
experimental data from \cite{Neill16}, shown as solid curves with markers for $\kappa_0$= $0.5$ and $2.5$.  
Numerical results are also plotted as dashed curves in Fig.~\ref{compk1tf1},
but are naturally indistinct from the respective analytical results.
More extensive analytical results have already been displayed in Figs.~(\ref{fig:entropy3qubit1})-~(\ref{fig:entropy3qubit2}) and discussed in the previous section.

The Fig.~(\ref{compk1tf1}(a,b)), corresponds to the initial state $|000\kt$, whose classical limit is the period-4 orbit.This classical the period-4 orbit is unstable at $\kappa_0=2.5$ and we see a rapid growth in the entanglement. However even at $\kappa_0=0.5$ entanglement grows to near maximal values, consistent with the large time average
 in Eq.~(\ref{eq:avgpiby2_1}), and with the analysis that predicts that the maximum occurs at a time scale $n_*\sim 3 \pi/\kappa_0 \sim 19$. 
We also noted that the entanglement at time $2n$ is same as the entanglement at time $2n-1$, see Eqs.~(\ref{eq:entsteps}) and ~(\ref{eq:concsteps}). Interestingly these are quite remarkably present (but previously unnoticed) in the experimental data for the first few time steps. All entanglement properties including concurrence
is left approximately unchanged in the experimental data for an odd-to-even time step as seen in Fig.~(\ref{compk1tf1}(a,b)). The degradation of this phenomenon is naturally to be attributed to decoherence and maybe a good measure of it.

The plots showing comparison of linear entropy and concurrence from the experimental data for the state $|\pi/2,-\pi/2\kt$ when $\kappa_0=0.5$ 
and $\kappa_0=2.5$, are shown in Fig.~(\ref{compk1tf1}(c,d)).
It shows a much smaller entropy growth for $\kappa_0=0.5$ in comparison 
to the state $|000\kt$, consistent with the bound in Eq.~(\ref{eq:pppbound}) and is a 
reflection, in the semiclassical limit, of the stable neighborhood of $|\pi/2,-\pi/2\kt$. This is also consistent with the long time
average, already displayed in Eq.~(\ref{eq:avgpiby2_2}).
More qualitative discussions of the time-evolution have been published in 
\cite{Madhok2018_corr}.


 \section{Exact solution for four-qubits:} 
\begin{figure}
 \centering
 \includegraphics[scale=1,keepaspectratio=true]{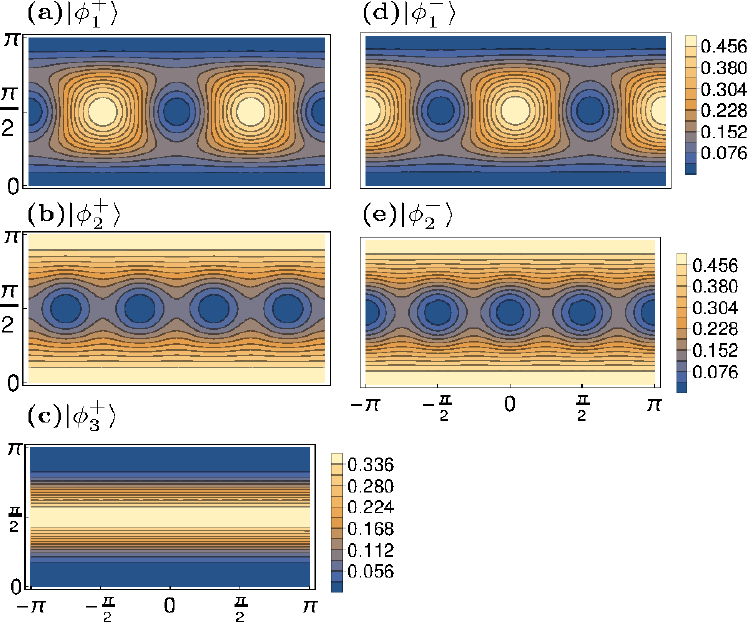}
 \caption{Husimi (quasiprobability distribution, $|\langle \phi_i|\theta_0,\phi_0 \rangle|^2$) plots
 for a set of five four-qubit bases states ($|\phi_i\rangle$), where $|\theta_0,\phi_0\rangle$ is 
 an arbitrary four-qubit, parametrized by ($\theta_0,\phi_0$).}
 \label{husimi4q}
\end{figure}

 It is particularly interesting to study a four-qubit
 kicked top as this is the smallest system where all-to-all interaction among qubits 
 is different from that of nearest-neighbour interaction, and therefore presents
 a special case of a genuinely nonintegrable system.  
 Surprisingly, even in this case, an exact solution to the kicked top with spin $j=2$, 
 is possible. 
 Similar to that of three-qubit kicked top, we are again confined to 
 ($2j+1=5$)-dimensional permutation symmetric subspace of the total $2^{2j}=16$-dimensional 
 Hilbert space.
 In this case the parity symmetry reduced and permutation symmetric basis
 in which $\mathcal{U}$ is block-diagonal is
 \begin{eqnarray}
  |\phi_1^{\pm} \rangle&=& \frac{1}{\sqrt{2}} (|W\rangle \mp | \overline{W} \rangle),\nonumber \\
 |\phi_2^{\pm} \rangle &=& \frac{1}{\sqrt{2}} (|0000\rangle \pm | 1111 \rangle), \nonumber \, \textrm{and}\\
 |\phi_3^{+} \rangle &=& \frac{1}{\sqrt{6}} \sum_{\mathcal{P}}|0011\rangle_{\mathcal{P}}
 \end{eqnarray}
where $|W\kt =\frac{1}{2}\sum_{\mathcal{P}}|0001\kt_{\mathcal{P}}$, $|\overline{W}\kt =\frac{1}{2}\sum_{\mathcal{P}}|1110\kt_{\mathcal{P}}$, and $\sum_{\mathcal{P}}$ sums over all possible permutations. Husimi plots for each of these states is shown in Fig.~(\ref{husimi4q}).
While all of these states $|\phi_j^{\pm}\kt$ are eigenstates of the parity operator $\otimes^4_{j=1} \sigma^y_j$ with eigenvalue $\pm 1$, a peculiarity of 4-qubits is that $|\phi_1^{+}\kt$ is also an eigenstate of the Floquet operator $\mathcal{U}$ with eigenvalue $-1$ for {\it all} values of the parameter $\kappa_0$. 

Thus the $5-$ dimensional space splits into $1\oplus2\oplus2$ subspaces on which the operators are $\mathcal{U}_0=-1$ and $\mathcal{U}_{\pm}$. 
Note that we continue to use the same symbol for the symmetry reduced Floquet operators as for the 3-qubit case, although they are not the same.
It is interesting that the eigenstate $|\phi_1^{+}\kt$ still has a classically viable interpretation, but only for small $\kappa_0$, where as 
is clear from the Husimi, it is localized on the fixed points and the symmetric islands. A more detailed study of eigenstates is postponed while we concentrate here on the time evolution.

In this basis, the unitary Floquet operator $\mathcal{U}$ becomes block diagonal, 
which makes it easy to take the $n^{th}$ power of the
unitary operator $ \mathcal{U}$,
\begin{equation}
\label{eq8}
  \mathcal{U}^n = \begin{pmatrix}
           (-1)^n & 0 & 0 \\ 0 &   
           \mathcal{U}_{+}^n & 0 \\ 0 & 0 &  \mathcal{U}_{-}^n
            \end{pmatrix}.
\end{equation}
Thus in this case also we do not encounter the need to take powers of any matrix other than 
$2$-dimensional ones.
Block $\mathcal{U}_{+}$ is $\mathcal{U}$ in the basis $\{\phi_2^{+},\phi_3^{+}\}$ and is,
\begin{equation}
\label{eq12}
  \mathcal{U}_{+} = -ie^{-\frac{i \kappa }{2}} \left(
\begin{array}{cc}
 \frac{i}{2} e^{-i \kappa} & \frac{\sqrt{3}i}{2}  e^{-i \kappa} \\
 \frac{\sqrt{3}i}{2}  e^{i \kappa} & -\frac{i}{2} e^{i \kappa} \\
\end{array}
\right),
\end{equation}
while $\mathcal{U}_{-}$ is $\mathcal{U}$ in the basis $\{\phi_1^{-},\phi_2^{-}\}$,
\begin{equation}
\label{eq12}
  \mathcal{U}_{-} = e^{-\frac{3 i \kappa }{4}} \left(
\begin{array}{cc}
 0 & e^{\frac{3 i \kappa }{4}} \\
 -e^{-\frac{3 i \kappa }{4}} & 0 \\
\end{array}
\right),
\end{equation}
where for simplicity we have used $\kappa=\kappa_0/2$.

Adopting the same procedure as for the case of 3 qubits, namely expressing $\mathcal{U}_+$ as a 
$SU(2)$ rotation, apart from a  phase, and taking its power results in 
\begin{eqnarray}
  \mathcal{U}_{+}^n &=&  e^{-\frac{i n(\pi+\kappa) }{2}} 
  \begin{pmatrix}
   \alpha_n & i\beta_n^{*} \\  i\beta_n & \alpha_n^{*}
  \end{pmatrix},
 \end{eqnarray}
where
 \begin{equation}
\alpha_n =  T_{n}(\chi)+\frac{i}{2}U_{n-1}(\chi)\cos{\kappa},\,
\beta_n  =  \frac{\sqrt{3}}{2}U_{n-1}(\chi)e^{i\kappa}.
 \end{equation}
As above, $T_{n}(\chi)$ and $U_{n-1}(\chi)$ denote the Chebyshev polynomials of the first 
and second kinds respectively, but now $\chi=\sin{\kappa}/2=\sin(\kappa_0/2)/2$. 

Similarly,
\begin{equation}
\label{eq12}
  \mathcal{U}_{-}^n = e^{-\frac{ 3in \kappa }{4}} \left(
\begin{array}{cc}
 \cos \frac{n\pi}{2} & e^{\frac{3 i \kappa}{4}} \sin \frac{n\pi}{2} \\
 -e^{-\frac{3 i \kappa }{4}} \sin \frac{n\pi}{2}  &  \cos \frac{n\pi}{2} \\
\end{array}
\right),
\end{equation}
which has a much simpler form than the $\mathcal{U}_{+}^n$ and in fact $\mathcal{U}_{-}^2=-e^{-3 i \kappa_0/4} I_2$ is up to a dynamical phase proportional
to the identity. Thus all states in the negative parity subspace are essentially periodic with period-2, a uniquely quantum feature. In particular the 
GHZ state $|\phi_2^{-}\kt=(|0000\kt -|1111\kt)/\sqrt{2}$ would be of this kind. 
Using these it is possible to find the exact evolution of the entanglement entropy of any one-qubit and again in particular we again concentrate on the initial states being $|0000\kt$ and $|\pm\pm\pm\pm \kt_y$, for the same reasons as in the 3 qubit case.

\subsection{Initial state $|\psi_0 \rangle=|0000\rangle$}
Considering four qubit state $|0000\rangle$, under the `$n$'
implementations of unitary operator $\mathcal{U}$,
\begin{eqnarray}
  \mathcal{U}^n |0000\rangle 
   &=& \frac{1}{\sqrt{2}} \left( \mathcal{U}^n_{+} |\phi_2^{+}\rangle
+ \mathcal{U}^n_{-} |\phi_2^{-}\rangle \right).
\end{eqnarray}
leads to the state $|\psi_n\rangle$ at time $n$. Just as for the 3 qubit case
the state $\mathcal{U}^{2n}|0000\kt$ is upto local-unitary operators same as
$\mathcal{U}^{2n-1}|0000\kt$, and therefore again all entanglement properties have 
``steps" in their dynamical evolution and it is sufficient to consider the time
$n$ to be an even integer. In this case
\begin{equation}
 |\psi_n\rangle =   e^{-\frac{i n(\pi+\kappa)}{2}} \frac{1}{\sqrt{2}} \left(
 \alpha_n |\phi_2^{+}\kt +i\beta_n |\phi_3^{+}\kt + e^{-\frac{in\kappa}{4}} |\phi_2^{-}\kt\right).
\end{equation}

\begin{figure}
 \centering
 \includegraphics[scale=1]{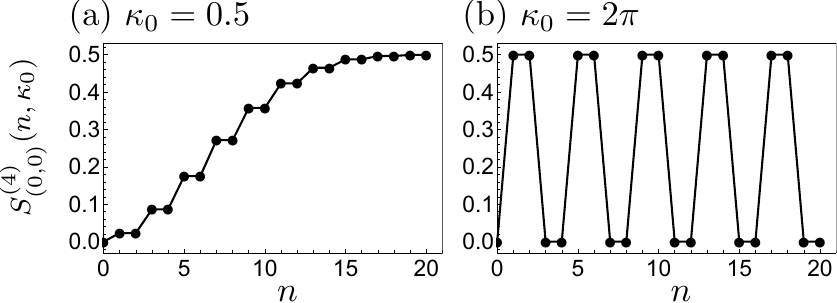}
 \caption{Linear entropy of a single-qubit reduced state versus $n$ is plotted
 at different values of $\kappa_0$ shown in parts (a) and (b), 
 corresponding to a four qubit initial state, $|0000\kt$.}
 \label{fig:entropy4qubit1}
\end{figure}

Single qubit reduced  density matrix is simply diagonal for even values of $n$,
eigenvalues being $\lambda(n,\kappa_0)$ and $1-\lambda(n,\kappa_0)$, 
where $\lambda(n,\kappa_0)=\frac{1}{2}\left( 1+ \xi_n(\kappa_0)\right)$,
where
\begin{eqnarray}
\label{eq:xi4qub}
 \xi_n(\kappa_0) &=& \textrm{Re} \left( \alpha_n e^{in\kappa_0/8} \right) \nonumber \\
 &=& T_{n}(\chi) \cos\frac{n\kappa_0}{8}-\frac{1}{2}U_{n-1}(\chi)\cos\frac{\kappa_0}{2} \sin\frac{n\kappa_0}{8}. \nonumber \\
\end{eqnarray}
For even values of $n$, linear entropy of a single-qubit 
reduced state is given by,
\begin{equation} \label{eq:ent4q1}
  S_{(0,0)}^{(4)}(n,\kappa_0)=\frac{1}{2}\left[ 1- \xi_n^2(\kappa_0) \right],
\end{equation}
and at odd values, $S_{(0,0)}^{(4)}(2n-1,\kappa_0)=S_{(0,0)}^{(4)}(2n,\kappa_0)$.
Figure~(\ref{fig:entropy4qubit1}) shows the evolution of this entanglement entropy for a few representative
values of $\kappa_0$.
In particular, even for $n=2$ (which is the same as $n=1$), we get a fairly long expression
for the entanglement entropy, hence rather than display it, we state that for small $\kappa_0$
it increases as $S_{(0,0)}^{(4)}(1,\kappa_0) \approx 3\, \kappa_0^2/32$, which is very similar
to the corresponding 3-qubit case. It grows monotonically with $\kappa_0$ till $\kappa_0=\pi$
where it attains the upper-bound of $1/2$ already, in contrast to the 3-qubit case which attains
this only at $\kappa_0=3 \pi/2$.

{To find the relevant time scales in the growth of the entanglement, we note that the maximum value
of the entropy is attained when $\xi_n(\kappa_0)=0$. From Eq.~(\ref{eq:xi4qub}), and noting that the zeros of $T_n(\chi)$ and $U_{n-1}(\chi)$,
do not occur simultaneously, we first examine the case when $n$ is even and $U_{n-1}(\chi)$ vanishes. This is similar to the analysis
of the $3$ qubit case above and we simply state that this implies that 
$n_* \approx 4 \pi/\kappa_0$. Thus the first even-time at which the second half of Eq.~(\ref{eq:xi4qub}) vanishes is $n_*$, however 
if this condition is satisfied the first part also vanishes as the $\cos( n\kappa_0/8)$ does. Thus the typical time-scale for 
the large entanglement to develop is slightly larger than the case of $3$ qubits where it was $3 \pi/\kappa_0$. At the time when the 
entanglement is maximum, $\beta_n \approx 0$ and the resultant states are superpositions of $|\phi_2 ^{\pm}\kt$ and are GHZ states.
Thus the large 1-qubit entanglement observed in the experiment of \cite{Neill16} for $\kappa_0=0.5$ has more to do with the creation of such 
GHZ states than thermalization or chaos.}

\begin{figure}[h]
 \includegraphics[width=8cm,keepaspectratio=true]{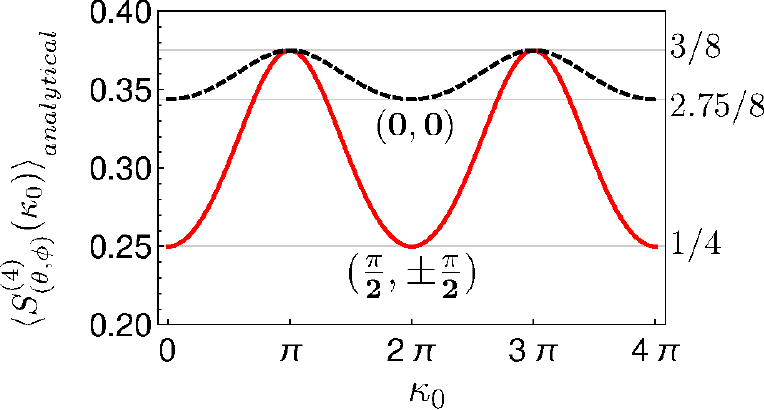}
 \caption{Analytically obtained 
expressions for time-averaged linear entropy for initial states $|0000\kt$ (Eq.~(\ref{eq:ent1})) 
and $|++++\kt_y$ (Eq.~(\ref{eq:ent2})) are plotted for $\kappa_0 \in (0,4\pi)$.
Extreme values are presented as horizontal lines, with their 
respective values ($S^{(4)}_{(\theta,\phi)}(\kappa_0)=\textrm{constant}$)
specified on the right side. Solid red curve and dashed black curve
correspond to initial states
$|++++\kt_y$ and $|0000\kt$ respectively.}
\label{fig:ana}
\end{figure}

Long time average of the linear entropy is obtained by averaging over the time $n$, and is given by 
\begin{eqnarray} \label{eq:ent1}
 \br S^{(4)}_{(0,0)}(\kappa_0)\kt &=&\frac{1}{8}\left( \frac{9+2 \cos^2(\kappa_0/2)}{3+\cos^2(\kappa_0/2)}\right), \kappa_0 \neq 0, 2 \pi.
\end{eqnarray}
For $\kappa_0=0$, or  $2 \pi$, the entanglement vanishes. As soon as 
$\kappa_0$ becomes non-zero, this long-time averaged linear entropy attains a
value of $2.75/8$, which further increases with $\kappa_0$ and 
attains a maximum value of $3/8$ at $\kappa_0=\pi$, as shown by the 
dashed curve in Fig.~(\ref{fig:ana}). 

Thus, in this case, long time-averaged 
linear entropy of single-qubit reduced state oscillates within a very small
interval of range $1/32$ for $\kappa_0 \in (0,2\pi)$.
\subsection{Initial state $|\psi_0 \rangle=|++++\rangle_y$}
\begin{figure}
 \centering
 \includegraphics[scale=1]{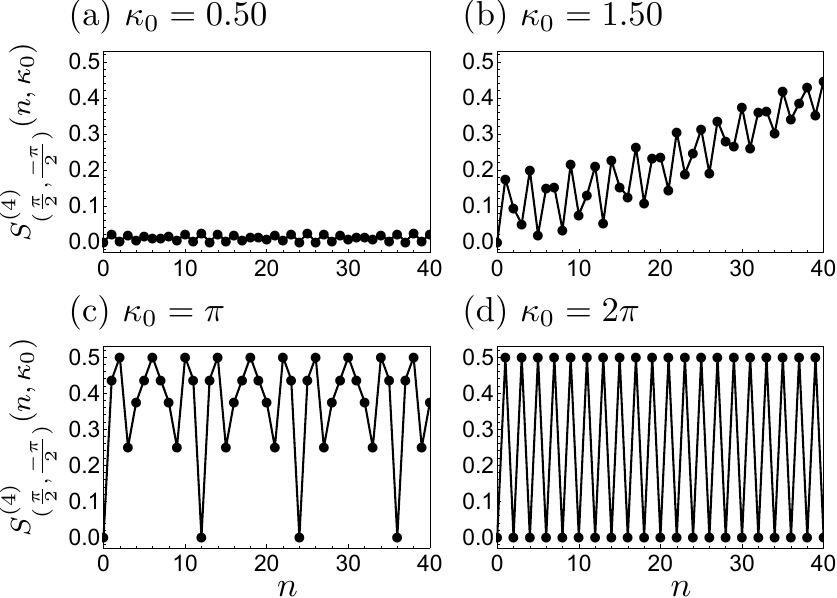}
 \caption{Linear entropy of a single-qubit reduced state versus $n$ is plotted
 at different values of $\kappa_0$ shown in parts (a), (b), (c), and (d), 
 corresponding to a four qubit initial state, $|++++\kt_y$.}
 \label{fig:entropy4qubit2}
\end{figure}

This state lies entirely in the positive parity subspace of 
the five dimensional permutation symmetric space of four qubits,
and is given by
\begin{equation}
\otimes^4 |+\kt_y = \frac{i}{\sqrt{2}}|\phi^+_1\kt + \frac{1}{\sqrt{8}} |\phi_2^{+}\rangle - \sqrt{\frac{3}{8}}|\phi_3^{+}\rangle,
\end{equation}
which under the action of $\mathcal{U}^n$, leads to $|\psi_n\rangle=\mathcal{U_{+}}^n|++++\rangle_y$, such that, (for $n>1$),
\begin{widetext}
\begin{align}
 |\psi_n\rangle = \frac{(-1)^{n}}{\sqrt{2}} \left( i |\phi_1^{+} \rangle 
+ e^{i \delta} \left( \alpha_n/2 - i \sqrt{3} \beta_n^{*}/2 \right) |\phi_2^{+} \rangle - e^{i \delta} \left( \sqrt{3} \alpha_n^{*}/2 - i \beta_n/2 \right) |\phi_3^{+} \rangle  \right),
\end{align}
\end{widetext}

where $\delta=n(2\pi-\kappa_0)/4$.
The reduced density matrix of any one of the four qubits is given by,
\begin{equation}
 \rho_1(n, \kappa_0)=\textrm{Tr}_{2,3,4}\left( |\psi_n\rangle\langle \psi_n | \right) = 
 \begin{pmatrix}
  1/2 & \xi'_n(\kappa_0) \\ \xi'_n(\kappa_0)^{*} & 1/2
 \end{pmatrix},
\end{equation}
where, \[ \xi'_n(\kappa_0) = - i(T_n(\chi) \cos \delta + U_{n-1}(\chi) \sin \delta \cos(\kappa_0/2)),\]
and the linear entropy is given by
\[ S^{(4)}_{(\pi/2,-\pi/2)}(n,\kappa_0)=  \frac{1}{2}\left( 1  - |\xi'_n(\kappa_0)|^2 \right). \]
Figure~(\ref{fig:entropy4qubit2}) shows the evolution of this entanglement entropy for a few representative
values of $\kappa_0$. 

A closed form expression for long time average linear entropy is then obtained as for
the other case and results in (for $\kappa_0 \neq 0, 2 \pi$),
\begin{equation} 
\label{eq:ent2}
 \br S^{(4)}_{(\frac{\pi}{2},\pm\frac{\pi}{2})} \kt  =\frac{1}{8}\left(\frac{9-\cos^2(\kappa_0/2)}{3+\cos^2(\kappa_0/2)}\right).
\end{equation}
As soon as $\kappa_0$ becomes non-zero, this long time-averaged linear entropy attains its
minimum value of $1/4$, which further increases with $\kappa_0$ and 
attains a maximum value of $3/8$ at $\kappa_0=\pi$, as shown by solid red
curve in Fig.~(\ref{fig:ana}). In this case, long time-averaged 
linear entropy of single-qubit reduced state oscillates within a relatively
larger interval of range $1/8$ for $\kappa_0 \in (0,4\pi)$.
\par
\begin{figure}[h]
 \includegraphics{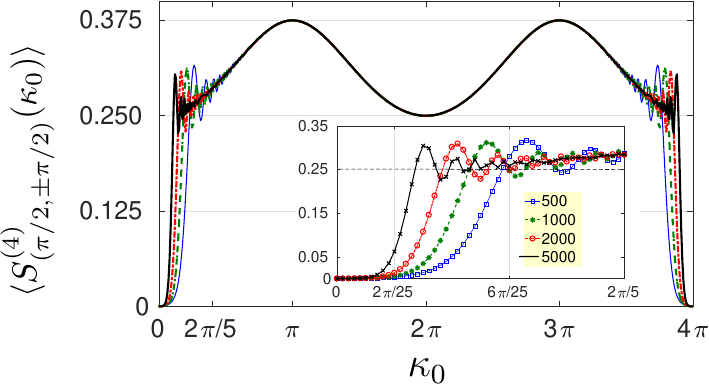}
 \caption{Simulated time-average linear entropy ($\br S^{(4)}_{(\frac{\pi}{2},\pm \frac{\pi}{2})}(\kappa_0)\kt$)
 subject to initial state, $|(\pi/2,\pm\pi/2)\kt$, plotted versus $\kappa_0$,
 is shown for different values of $n$ (as given in the inset) in the interval $\kappa_0 \in [0,4\pi]$.
 Inset shows the blowed-up horizontal scale for $\kappa_0 \in [0,\Delta \kappa_0]$, 
 where $\Delta \kappa_0=2\pi/5$, that clearly presents the curves approaching the 
 solid curve curve of Fig.~\ref{fig:ana}.}
\label{fig:savg4q}
\end{figure}

Time averaged linear entropy of single-qubit reduced state in both of these cases, reach their maximum value of $3/8$ when $\kappa_0=\pi$ and, remarkably, this matches with the average from the ensemble of random permutation symmetric states \cite{SheshadriPRE2018} of 4-qubits $S_{RMT}(4)$ as in the case of the 3-qubit case. In addition we  see that the average for the states at $(\pi/2, \pm \pi/2)$ attain the value of $1/4$ for arbitrarily small $\kappa_0$ in contrast to the 3-qubit case which vanishes as in Eq.~(\ref{eq:avgpiby2_2}). In fact the non-zero average is seen in numerical calculations to be attained only on averaging over extremely long times for small $\kappa_0$, that 
reflects in Fig.~(\ref{fig:savg4q}), where different curves correspond to time-average over different times (as labelled 
in terms of $n$ in the inset). 
For small values of $\kappa_0$, i.e. $\kappa_0=2p\pi\pm \Delta \kappa_0$ ($p$ being an integer),
time-average behavior of linear entropy for different times does not converge, 
and approaches the infinite-time average consistent with Eq.~(\ref{eq:ent2}) and Fig.~(\ref{fig:ana}),
as $n\rightarrow \infty$.
This slow thermalization, specifically for state $|(\pi/2,\pm\pi/2)\kt$ is 
attributed to the process of dynamical tunneling to which we now turn.
\subsection{Dynamical tunneling}
\begin{figure}[h]
 \includegraphics{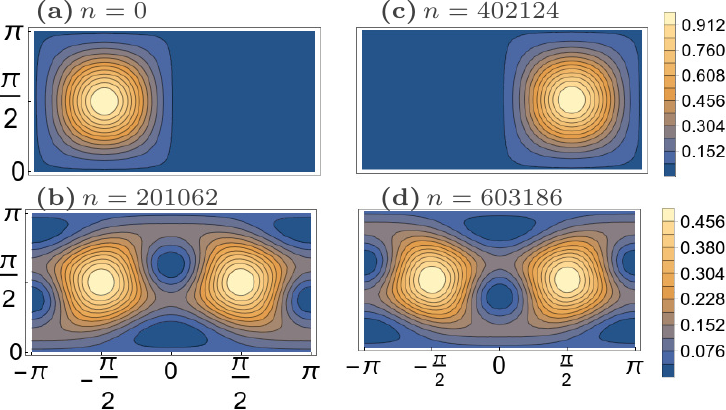}
 \caption{Husimi (quasi probability distribution) plots for the four-qubit initial state,  $\otimes^4|+\kt$, evolving under $n$ implementations of 
 $\mathcal{U}$, and leading to tunneling to the state,  $\otimes^4|-\kt$, at time $n_* \approx 128 \pi/\kappa_0^3 \approx 402124$. ($\kappa_0=0.1$). }
\label{fig:tunnel}
\end{figure}

This very slow process is due to tunneling between $\otimes^4|+\kt_y$ and $\otimes^4|-\kt_y$. At $\kappa_0=0$, two positive parity eigenvectors of $\mathcal{U}$,  $|\phi_1^+\kt$ and $|\phi_{23}^+\kt=\frac{1}{2} |\phi_2^{+}\rangle - \frac{\sqrt{3}}{2}|\phi_3^{+}\rangle$ are degenerate with eigenvalue $-1$. 
These can also be written as 4-qubit GHZ states \cite{GHZ0,GHZ}: 
\begin{equation}
 i |\phi_1^+\kt=\left(\otimes^4|+\kt_y -\otimes^4|-\kt_y \right)/\sqrt{2},
\end{equation}
the unchanging eigenstate,  and 
\begin{equation}
 |\phi_{23}\kt = \left(\otimes^4|+\kt_y +\otimes^4|-\kt_y \right)/\sqrt{2}.
\end{equation}
Thus
\begin{equation}
\mathcal{U}^n\otimes^4 |+\kt _y = (-1)^{n}\frac{i}{\sqrt{2}}|\phi_1^+\kt + \mathcal{U}_+^n \frac{1}{\sqrt{2}}|\phi_{23}^+\kt.
\label{eq:tunnelevolve}
\end{equation}
The eigenvalue of $\mathcal{U}_+$ that is $-1$ at $\kappa_0=0$ is $e^{i \gamma_{-}}$ with 
\begin{equation}
\gamma_{-}= \frac{\kappa_0}{4}+\pi -\sin^{-1}\left(\frac{1}{2}\sin \frac{\kappa_0}{2} \right) \approx \pi -\frac{\kappa_0^3}{128}.
\end{equation}
This implies that for $\kappa_0 \ll 1$, the corresponding state and $|\phi_1^+\kt$ are nearly degenerate. The splitting leads to a change in the relative phase of their contributions in Eq.~(\ref{eq:tunnelevolve})
and at time $n_* \approx 128 \pi/\kappa_0^3$ the evolved state is close to $\otimes^4|-\kt$, leading to tunneling as shown in Fig.~(\ref{fig:tunnel}) between what in the classical limit are two stable islands. At time $n=n_*/2$ the state obtained is close to the GHZ state $(\otimes^4 |+\kt_y -i \otimes^4 |-\kt_y)/\sqrt{2}$. 

This tunneling is observed whenever $\otimes^{2j}|\pm\kt$ are degenerate eigenstates of the rotation part of the Floquet $\mathcal{U}$. This implies that the number of qubits should be an integer multiple of $2\pi/p$, where
$p$ is the rotation angle (we have used $p=\pi/2$, and hence the tunneling occurs when the number of qubits is a multiple of 4).

\begin{figure}[h]
 \includegraphics{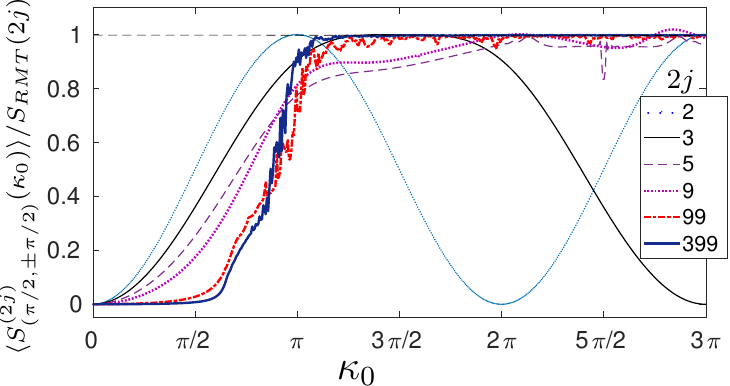}
 \caption{Normalized average single-qubit entanglement when the initial state is $\otimes^{2j}|+\kt_y$ for increasing number of qubits
 (except multiples of $4$ where there is tunneling for $p = \pi/2$.)}
\label{fig:avgentlargerj}
\end{figure}
For larger number of qubits, the average single-qubit entropy, normalized by the random state average,
is numerically found when the initial state is $\otimes^{2j}|+\kt_y$ and shown in Fig.~(\ref{fig:avgentlargerj}). The number of qubits used
for the cases shown in this figure are not multiples of $4$, and hence
the long-time average vanishes at $\kappa_0=0$, that is there is no tunneling. The trend is in keeping with a more complex classical phase space that becomes fully chaotic when the random state average is approached. The initial state being centered on a fixed point, increasing the number of qubits leads to a sharp growth beyond $\kappa_0=2$ when the fixed point becomes unstable, a more detailed  study of this is found in \cite{Bhosale-pre-2017}, without the connection to tunneling. Interestingly even for the 3-qubit case, for which we have the analytical evaluation in Eqs~(\ref{eq:avgpiby2_2}), a similar but smoother trend is displayed and reaches the random state value when $\kappa_0=3\pi/2$.
\begin{figure}[h]
\begin{center}
 \includegraphics{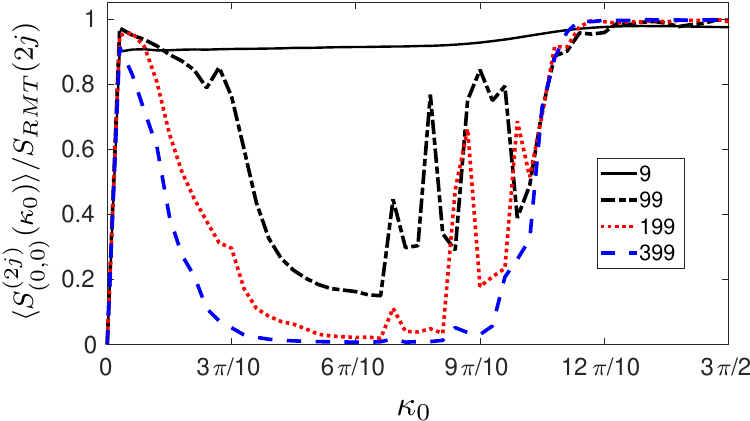}
\end{center}
 \caption{Normalized average single-qubit entanglement when the initial state is $\otimes^{2j}|0\kt$ for increasing number of qubits.}
\label{fig:savgall0largej}
\end{figure} 

{The complementary state $\otimes_{k=1}^{2j}|0\kt$ has a nonzero average 
as $\kappa_0$ approaches $0$, both for the 3- and the 4-qubit cases. We have already discussed the origin of this in some detail for the $3-$ qubit case. For a very large number of qubits we expect that classically the tunneling effect vanishes. This is borne out in Fig.~(\ref{fig:savgall0largej}), although surprisingly even for very large number of qubits for a range of $\kappa_0$ values close to $0$,
the formation of nonclassical states resulting in large average single qubit entanglement is seen. The subsequent increase of entanglement for larger values of $\kappa_0$ is due to the destabilization of the period-4 orbit at $\kappa_0=\pi$. A more detailed analysis is called for, including the study of entanglement between large blocks of spins which will distinguish between the non-classical states produced when the system is near-integrable and the random states produced at much larger values of the parameter when the classical phase-space is mixed or chaotic. A recent analysis in \cite{kumari2018untangling}
uses an upper-bound of the entanglement entropy using the Fannes-Audernaert inequality to argue for connections between entanglement and chaos and why 
states localized on the stable period-4 orbits can have large entanglement in the deep quantum regime.

 \section{Conclusions}

Quest for an exactly solvable model is hard and often a matter of serendipity. In our work, we give exact solution for 3- and 4- qubit instances of the kicked top
and explicitly derive expressions for the time evolved state, reduced density matrix, entanglement entropy and its long time average values. 
Our work provides interesting connections between a quantum system with few degrees of freedom and its classical limit that is non-integrable and can exhibit chaos for high $\kappa_0$ values.
 For example, we find that the exactly solvable 3- and 4- qubit instances of the kicked top provide insights
into how entropy and entanglement thermalize in closed quantum systems in the sense of long time averages approaching ensemble
averages, as the classical limit approaches global chaos, as predicted by random matrix theory. Since we derive exact analytical results valid for all values of $\kappa_0$, this will be further useful to study transition to thermalisation in closed quantum systems. Experiments have already probed the 3-qubit case, and it is worth mentioning that, in the light of our work, it should now be viewed as a study of thermalisation in an integrable system rather than thermalisation induced due to lack of sufficient number of conserved quantities \cite{Neill16}. It will be interesting to see at what spin size does the exact solvability of these models become intractable and whether or not that has a physical interpretation.

Even more remarkable is the entanglement dynamics at small values of the chaoticity parameter. This cannot be directly attributed to non-integrability.
For example, even for small $\kappa_0$, in the case of the three qubit  $|0,0\kt$ state, we find an increase of entanglement with time, which can be attributed to the generation of highly non-classical GHZ type states. We accurately predict time scales for such entanglement dynamics and
found an excellent agreement with the numerics. Likewise, the  $|\pi/2, -\pi/2\kt$ state in the 4-qubit case displays, for the same rotation angle, tunneling and creation of GHZ states and we have described this in detail as well. In the near-integrable regime we exactly calculate tunneling splitting and show this to be in  agreement with the numerics. To the best of our knowledge, ours is the first work to find a connection between tunneling splitting, the number of qubits 
and a system parameter.
It is worth mentioning that entanglement generation occurs despite the initial state being localised on a stable island with phase space having almost no chaos.
We believe our findings and analysis of entanglement generation at low values of  $\kappa_0$ will contribute to the understanding of entanglement generation
in dynamical systems and it's connections to classical bifurcations, emergence of structures and ergodicity in the phase space. This also complements our findings for higher values of $\kappa_0$
as well as the existing literature on the connections between entanglement generation and chaos.

Lastly, larger number of qubits can show genuine signatures of non-integrability and chaos, and tunneling leads to creation of macroscopic superpositions that are generalized GHZ states. We hope our work raises new questions and adds to the discussion on the connections between integrability, quantum chaos, and thermalization. Since the multi-qubit kicked top can be viewed as an analog quantum simulator, robustness of such a system to errors \cite{hauke2012can, Shepelyansky_2001}, especially in the regime where we generate highly non-classical GHZ like states and explore truly quantum phenomena like tunneling, will be of interest to the quantum information community. As an aside, we are able to give an alternate proof of the Pell identity satisfied by the Chebyshev polynomials!

\begin{acknowledgments}
We are grateful to the authors of \cite{Neill16} for generously sharing their experimental data, in particular to Pedram Roushan and Charles Neill for useful correspondence regarding the same.
\end{acknowledgments}  
\appendix*
\section{ Linear entropy of an arbitrary three-qubit 
  permutation symmetric state\label{app3}}
\begin{figure}[h]
\includegraphics[scale=0.9]{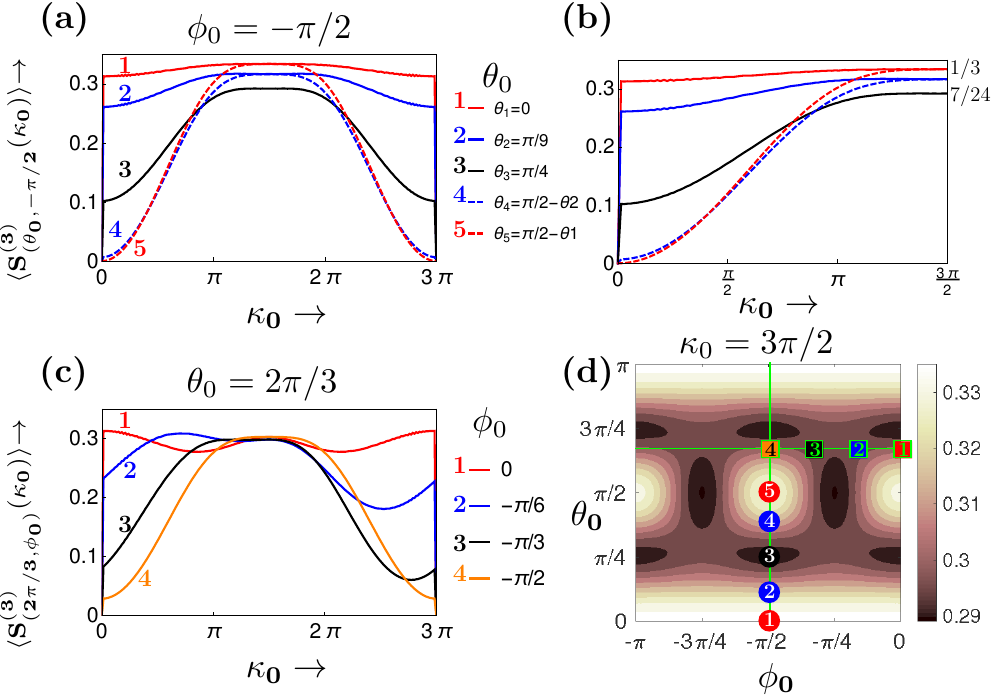}
\caption{(a),(b) Time averaged linear entropy ($\langle  S_{(\theta_0,-\pi/2)}^{(3)} \rangle$)
 of a single party reduced state vs chaoticity parameter $\kappa_0$. 
 Different curves correspond to different
 initial states, $|\theta_0,-\pi/2\rangle$ as labelled $1$ to $5$, 
 alongwith explicit values of $\theta_0$ given in the plot legends.
 These corresponding initial states $|\theta_0,\phi_0\kt$ are also 
 marked as numbered circles in the contour plot given in part (d).
 Part (c) contains the plots for $\langle  S_{(2\pi/3,\phi_0)}^{(3)} \rangle$
 vs chaoticity parameter $\kappa_0$ for a fixed value of $\theta_0=2\pi/3$. 
 Different curves correspond to different
 initial states, labelled by numbers $1$ to $4$
 alongwith explicit values of $\phi_0$ given in the plot legends.
 Respective initial states $|\theta_0,\phi_0\kt$ are also 
 marked as numbered squares (with a green border) in the contour
 plot given in part (d). Contour plot shown in part (d) corresponds to 
 $\kappa_0=3\pi/2$. \label{entropyplot}}
\end{figure}

Considering a three-qubit state
\begin{equation}
\label{eqg1n}
 |\psi_0\rangle= 
 a_1 |\phi_1^{+}\rangle + a_2 |\phi_2^{+}\rangle +
 b_1 |\phi_1^{-}\rangle + b_2 |\phi_2^{-}\rangle.
\end{equation}
Each of the three qubits are initialized in the same state 
($|\psi \kt=\cos \frac{\theta_0}{2} |0\kt + e^{-i \phi_0} \sin \frac{\theta_0}{2} |1\kt$, 
in the computational bases), such that the initial state of the
$3-$ qubit system is $|\psi_0 \kt=\otimes^{3} |\psi \kt$,
where $\theta_0 \in [0,\pi]$ and $\phi_0 \in [-\pi, \pi]$.
Repeated implementations of the unitary operator $\mathcal{U}$, 
leads to $|\psi_n\kt=\mathcal{U}^n|\psi_0\kt$.
We obtain single-party reduced density operator
 by tracing out any of the two qubits of 
the three-qubit density operator ($\rho_n=|\psi_n\kt \langle \psi_n|$),
leading to,
\begin{equation} \label{eqa1}
 \rho_i = \begin{pmatrix}
             r & s \\ s^{*} & 1-r 
             \end{pmatrix},
\end{equation}
where the elements of the density operator are
given by
\begin{eqnarray}
 r &=& \frac{1}{2}+\textrm{Re}\left( a_{1n}b_{1n}^{*} + \frac{1}{3}a_{2n}b_{2n}^{*} \right)  
 \quad \textrm{and}\nonumber \\ 
 s &=& \frac{1}{\sqrt{3}} \textrm{Re}\left(a_{1n}b_{2n}^{*} + b_{1n}a_{2n}^{*} \right) 
 + \frac{i}{\sqrt{3}} \textrm{Im}\left( a_{1n}a_{2n}^{*} + b_{1n}b_{2n}^{*} \right) \nonumber \\ 
 & & -\frac{i}{3} \left( a_{2n} + b_{2n} \right) \left( a_{2n}^{*} - b_{2n}^{*} \right).
\end{eqnarray}
Where the coefficients, $a_{1n}=a_1  \alpha_n -  a_2 \beta_n^{*}$, 
$a_{2n}=a_1 \beta_n +  a_2 \alpha_n^{*}$, $b_{1n}=i^{n} \left(
 b_1  \alpha_n + b_2 \beta_n^{*} \right)$, and $b_{2n}=i^{n} \left(
 b_2  \alpha_n^{*} - b_1 \beta_n \right)$.
Linear entropy of the single-qubit (Eq.~(\ref{eqa1})) is thus given by,
\begin{equation} \label{eqa2}
 S_{(\theta_0,\phi_0)}^{(3)}(n,\kappa)=2 \left[ r(1-r)-|s|^2 \right].
\end{equation}
Thus linear entropy is obtained as a function of the initial-state parameters ($\theta_0, \phi_0$).
Long time average linear entropy is calculated numerically with $n=1000$
for various initial states as shown in Fig.~(\ref{entropyplot}).
Part (a) and (c) of Fig.~(\ref{entropyplot}) show the variation of 
time average entropy with chaoticity parameter for a period $2\pi j$.
Pairs of complimentary $\theta_0$s, saturate to same values 
in the region around $\kappa_0=3\pi/2$. 
Part (b) of Fig.~(\ref{entropyplot}) highlights the range of values of 
average linear entropy at $\kappa_0=3\pi/2$, a scale of similar range 
in part (d) depicts that the linear entropy of a single-qubit reduced 
state for an arbitrary value of parameters ($\theta_0, \phi_0$) fall 
into this range. 
Further, we have obtained an explicit closed form experssion for 
long time average linear entropy for an arbitrary ($\theta_0, \phi_0$)
at $\kappa_0=3\pi/2$, which is discussed in the main text.

\end{document}